%% file: main.tex
\documentclass[fleqn,10pt]{wlscirep}
\usepackage[utf8]{inputenc}
\usepackage[T1]{fontenc}
\usepackage{multirow}

\usepackage{graphicx}
\usepackage{dcolumn}
\usepackage{bm}
\usepackage{amsmath}
\usepackage{amssymb}
\usepackage{tocloft}
\usepackage{xcolor}
\usepackage[colorlinks=true]{hyperref}
\usepackage{caption}
\usepackage{algpseudocode}
\usepackage{algorithm}
\usepackage[most]{tcolorbox}
\usepackage{enumitem}
\usepackage{braket}
\usepackage{float}
\usepackage[qm]{qcircuit}
\usepackage{tikz} 
\usepackage{pgfplots} \pgfplotsset{compat=1.18} 
\usepgfplotslibrary{patchplots}
\usetikzlibrary{decorations.pathreplacing, shapes, positioning}
\usepackage{subcaption}
\usepackage{booktabs}
\usepgfplotslibrary{fillbetween}

\newtcbtheorem[]{example}{Example}%
  {enhanced,
   breakable,
   colback=purple!10!white,
   colframe=purple!65!blue,
   fonttitle=\bfseries,
   listing only
  }{example}

\newtheorem{theorem}{Theorem}

\title{Quantum Decoding Algorithms: Quantum Speedups in Optimization}

\author[1,2]{Jan Ljubas}
\author[2,3,4,$*$]{Tim~Byrnes}

\affil[1]{University of Copenhagen, Department of Mathematical Sciences, Universitetsparken 5, Copenhagen, Denmark}
\affil[2]{New York University Shanghai; NYU-ECNU Institute of Physics at NYU Shanghai, 567 West Yangsi Road, Pudong, Shanghai 200126, China}
\affil[3]{Center for Quantum and Topological Systems (CQTS),
NYUAD Research Institute, New York University Abu Dhabi, UAE}
\affil[4]{Department of Physics, New York University, New York, NY, 10003, USA}
\affil[$*$]{tim.byrnes@nyu.edu}

\begin{abstract}
Attaining a quantum speedup in solving practically useful optimization problems has been one of the holy grails in the field of quantum computing. While prior approaches have demonstrated speedups for certain structured problem classes, establishing a clear and scalable advantage on broadly useful practical optimization problems remains challenging.  Recently, a new approach to solving the max-LINSAT class of optimization problems has emerged, called Decoded Quantum Interferometry (DQI).  In DQI, a combination of techniques rooted in (classical) coding theory and interferometry are used to obtain the solution of max-LINSAT.  In the special problem instance of the optimal polynomial intersection (OPI) problem, strong evidence exists to show that an superpolynomial speedup exists over the best classical methods in obtaining an approximate solution.  In this review, we give a self-contained description of DQI and the necessary background to understand the algorithm.  Specifically, we give the essentials of Galois fields, optimization problems such as max-LINSAT and OPI, and coding theory, followed by a step-by-step walkthrough of the quantum algorithm and its operating principle.    
\end{abstract}

\begin{document}

\flushbottom
\maketitle

\thispagestyle{empty}


   


\section{Introduction}

The quest for finding a quantum speedup to various computational problems has been a driving force in quantum computing since its earliest days \cite{montanaro2016quantum}.  The early discoveries of Shor's factoring algorithm \cite{Shor_1997} and Grover's search \cite{grover1996fastquantummechanicalalgorithm} showing exponential and quadratic speedups respectively, kicked off interest in quantum information science which continues to this day. 
Beyond showing an improvement in an algorithmic sense, demonstrating that a quantum computer has a considerable and tangible speedup --- including potential real-world effects such as decoherence and control errors --- is now an important aim, a concept called quantum advantage.  The Google collaboration showed that a quantum computer could perform a random circuit sampling task faster than the best available classical computer \cite{arute2019quantum}.  This was shortly followed by the demonstration of quantum advantage in optical circuits performing Gaussian boson sampling \cite{zhong2020quantum}.  While such sampling problems do not solve any useful real-world problem, they still demonstrate the potential for quantum computers to be superior in performance to classical computers. More recently, efforts have been directed to demonstrate quantum utility \cite{herrmann2023quantum}, where a quantum computer aims to solve a practically useful problem with superior performance compared to a classical computer. 

Optimization problems have been long conjectured to be a problem class that may be suitable for quantum computers to solve \cite{farhi2001quantum}.  They also have a broad practical interest, with applications in planning, logistics, manufacturing, financial portfolio management, computer vision, artificial intelligence, machine learning, bioinformatics, drug design, and a variety of chemical and physical materials problems \cite{tanahashi2019application, smelyanskiy2012near,hauke2020perspectives,mohseni2022ising}.   Early quantum optimization approaches, such as quantum annealing \cite{ray1989sherrington,farhi2001quantum,rajak2023quantum,das2008colloquium} showed promise but have yet to demonstrate broadly applicable, scalable quantum advantage on practically relevant optimization problems. Speedups that have been observed tend to be on structured or carefully constructed benchmark instances where quantum annealing is expected to perform favorably \cite{mohseni2022ising}.  Another related approach is the quantum approximate optimization algorithm (QAOA) \cite{farhi2014quantum}, which can be considered a discretized version of quantum annealing, where optimal parameters are determined by a 
 variational quantum-classical feedback loop.  Here, in certain structured problem families, a scaling advantage has been shown, such as the low autocorrelation binary sequences problem \cite{shaydulin2024evidence}. Other approaches such as Grover's algorithm and amplitude amplification applied to optimization problems generally provide only quadratic speedups over unstructured classical search, which is often insufficient to outperform the best specialized classical optimization algorithms for structured problems.  A quadratic speedup on an exponentially scaling problem is still exponential, and typically classical algorithms have superior scaling compared to a naive application of Grover's algorithm. Hence despite the intense activity in finding a quantum speedup for optimization problems, demonstrated quantum optimization advantages have largely been confined to specific structured problem classes, and the extent to which these results translate to broad practical real-world optimization tasks remains an open question. For more on provable and heuristic quantum speedups in optimization, see Refs. \cite{sanders2020compilation,Abbas2024,Quinton2025,Dalzell_2025,Egger2025quantumoptimization}.


Recently, a new approach to quantum optimization has attracted a great deal of attention, called Decoded Quantum Interferometry (DQI) \cite{jordan2024}. Using methods rooted in coding theory, it was shown that the DQI algorithm is capable of solving the max-LINSAT class of optimization problems. For particular subclasses of the max-LINSAT problem, such as the Optimal Polynomial Intersection (OPI) problem, strong evidence suggests that there is an superpolynomial quantum speedup in obtaining an approximate solution. The speedup largely depends upon whether the dual problem of the optimization problem can be efficiently solved, which relies upon methods from coding theory and cryptography. Therefore, while not every max-LINSAT problem can be solved efficiently, particular subclasses of it can be solved efficiently by DQI.  One of the interesting aspects of the method is that it works on a different operating principle to previous methods for solving optimization problems on a quantum computer, where one of the key steps is to use a quantum Fourier transform to construct the solution. Furthermore, the types of problem that it is capable of solving have a structure that appears to be potentially applicable to real-world problems. For example, the OPI problem is essentially a polynomial fitting problem.  While it is still early days since the discovery of the algorithm, it has generated much interest and numerous generalizations have already been proposed \cite{schmidhuber2025HDQI,sabater2025BMW_DQI,ralli2025QChem,bu2025DQIUnderNoise,khattar2025verifiableQuantumAdvantageDQI,BlanvillainChaillouxTillich2025_QDP,briaud2025quantumAdvantageMultivariatePolynomials}.  



The theory behind DQI stems from results in the early-2000s on quantum algorithms for lattice problems \cite{AharonovTaShma2003} \cite{AharonovRegev2005} and quantum complexity of lattice problems \cite{GuruswamiMicciancioRegev2005} \cite{micciancio_regev_2009} such as the Shortest Vector Problem (SVP), Short Integer Solution problem (SIS) and their numerous variants.  In the seminal papers of Refs. \cite{regev04,regev2005lwe}, motivated by reductions between dual problems in lattice-based cryptography and building quantum cryptosystems, Regev demonstrated how to reduce the SIS problem to the Learning With Errors (LWE) problem. 
Since then, further strides were made towards better understanding of quantum lattice complexity and the role of Regev's reduction \cite{BlanvillainChaillouxTillich2025_QDP, ChaillouxHermouet2025_S_LWE_ISIS, Chen2025}, see also Refs.  \cite{c_t_soft, CT_q_decoding2023, BlanvillainChaillouxTillich2025_QDP}. 
Apart from being a tool for proving hardness, Regev's reduction is a powerful quantum algorithmic primitive. One of the first notable algorithmic extensions of Regev's reduction was made by Chen, Liu and Zhandry \cite{ChenLiuZhandry2021}, which introduced new, filtering-based quantum algorithms for specific inhomogeneous linear formulations of the LWE problem.
This was followed by Yamakawa and Zhandry's breakthrough work \cite{YZ22}, which showed that a super-polynomial quantum speedup could be attained  using Regev's reduction. Their algorithms solve certain NP search problems by utilizing Regev's lattice-based reduction to transform the decoding of high-dimensional linear codes into a quantum-friendly framework. The significance of this work lies in achieving a quantum speedup without relying on structured oracles, such as those with periodicity, which have been central to previous results such as Shor's algorithm. 
Following similar ideas, Chailloux and Tillich explicitly formalized the quantum decoding problem \cite{CT_q_decoding2023}, essential in later theoretical work, and suggested quantum decoding algorithms with Regev’s reduction, matching the best known quantum algorithms at the time for the short codeword problem (similar to SIS).



In this review, we provide a simple introduction of the DQI algorithm and the associated background.   The goal is to bring theoretical framework closer to physicists, computer scientists, and engineers which we hope will help bridge the gap between theory and practical application of the DQI and related algorithms. More generally, we call the algorithms that follow this general approach {\it quantum decoding algorithms}, as the core ideas stem from the concept of reducing the optimization problems to a decoding problem. 
To make this review self-contained, we first give a brief introduction to Galois fields, which is the mathematical foundation underpinning coding theory and the optimization problems that we consider (Sec. \ref{sec:fields}). We then discuss the max-LINSAT problem and its special cases, which are optimization problems that are solved by DQI (Sec. \ref{sec:optimization}).  This is followed by a brief description of coding theory, which is the classical theory of error correction (Sec. \ref{sec:coding}).  Some understanding of coding theory is necessary in DQI as one of the crucial steps is a classical decoding step.  Section \ref{DQI_explanation} is the core content of this review and explains the DQI algorithm, walks through the algorithm step-by-step and discusses the nature of the quantum state that is prepared.  We discuss the performance of the DQI algorithm in Sec. \ref{sec:performance}, and give the extensions and further developments in Sec. \ref{DQI_extensions}.  Finally, in Section \ref{Discussion_future_work} we summarize and discuss open questions.



\section{Galois Fields} 
\label{sec:fields}

In mathematics, a \textit{finite (Galois) field $\mathbb{F}_{q}$} is a field  (closed under multiplication, addition and their inverses) with $q$ elements, where $q =p^l$ and $ p $ is a prime number with $ l \ge 1 $. 
For $ l = 1 $, it corresponds to mod-$p$ modular arithmetic, such that the field  $\mathbb{F}_p$ consists of the integer set $\{0, 1, ..., p-1\}$.  
In this review we will be mainly concerned with the $ l = 1 $ case, hence we take $ q = p $.  The \textit{order} $|\mathbb{F}_{p}|$ of a field $\mathbb{F}_{p}$ is the number of its elements, hence $|\mathbb{F}_{p}| = p$. 
We also define $\mathbb{F}_p^\times$ as the subset of all non-zero elements of $\mathbb{F}_p$.  

A \textit{primitive element $\gamma$} of a field $\mathbb{F}_p$ is a generator of the multiplicative group $\mathbb{F}_p^\times$, so that  $\{\gamma^0, \gamma^1, ..., \gamma^{p-2}\} = \mathbb{F}_p^\times$. In other words, a primitive element $\gamma \in \mathbb{F}_p $ is an element of that field whose powers generate all non-zero elements of the field, i.e. $\{\gamma^k : k \in [0, p-2] \} = \mathbb{F}_p^\times$. We say that $\gamma$ generates $\mathbb{F}_p^\times$. 
$\mathbb{F}_p$ is also denoted as $\mathbb{F}_p \cong  \mathbb{Z}/p\mathbb{Z}$. This notation hints the existence of a natural (canonical) ring homomorphism
\begin{align}
\pi(a) = a \text{ mod }p,
\end{align}
which is a surjective map which sends each integer to its equivalence class modulo $p$.

A polynomial $P \in \mathbb{F}_p [x] $ is defined as 
\begin{align}
\label{polynomial}
P(x) = \sum_{j=0}^{n-1} a_j x^j  = a_0 + a_1 x + \dots + a_{n-1}x^{n-1} , 
\end{align}
where all coefficients $a_j \in \mathbb{F}_p $. 
The polynomial $P$ defines a function $\mathbb{F}_p \rightarrow \mathbb{F}_p$, i.e. $x, P(x) \in \mathbb{F}_p $. Since both its domain and codomain are finite and discrete, a plot of $P(x)$ is not a curve, but a set of $p$ discrete points (an explicit example is given in Example \ref{example:OPI_instance}). As such, in contrast to the polynomials over real numbers, there is no notion of smoothness, continuity, or ordering.  


In addition, we can extend the above modular map $\pi(x)$ (coefficient-wise) to polynomials: 
\begin{align}
\pi ( P(x) ) = \sum_{j=0}^{n-1} \pi (a_j) x^j . 
\end{align}
This is a surjective ring homomorphism, and its kernel is the ideal $p\mathbb{Z}[x]$, i.e. all polynomials whose coefficients are multiples of $p$. What this means is that every polynomial $P(x)$ with \textit{integer} coefficients has a unique image $Q(x) = \pi ( P(x) )  \in \mathbb{F}_p[x] $, obtained by applying $\pi $ to all coefficients of $P(x)$. That reduction is well-defined and unique because each integer $a \in \mathbb{Z}$ has a unique residue class modulo $ p$  in $\mathbb{Z}/p\mathbb{Z}  \cong \mathbb{F}_p $.
Mapping polynomials with integer coefficients to polynomials over $\mathbb{F}_p$ may lead to information loss, as this map is not injective (multiple polynomials $P(x)$ map to the same polynomial $Q(x) \in \mathbb{F}_p[x] $).  An explicit example of the above are shown in Example \ref{example:F5_arithmetic}.

\begin{example}{Arithmetic in $\mathbb{F}_5$}{F5_arithmetic}
Consider the finite field $\mathbb{F}_5$. Since $ q = 5 $ is prime, we can take the elements to be $\{0, 1, 2, 3, 4\}$ and operations to be  performed modulo 5.
\begin{itemize}[noitemsep, topsep=0pt]
    \item \textbf{Addition:} Sums wrap around after reaching 4. For example, $3 + 4 = 7 \equiv 2 \pmod 5$.
    \item \textbf{Multiplication:} Products also wrap around. For example, $3 \cdot 4 = 12 \equiv 2 \pmod 5$.
    \item \textbf{Inverses:} Every non-zero element $a$ has a unique multiplicative inverse $a^{-1}$ such that $a \cdot a^{-1} \equiv 1 \pmod 5$ --- e.g. in $\mathbb{F}_5$, the inverse of $2$ is $3$, because $2 \cdot 3 = 6 \equiv 1 \pmod 5$. 
    \item \textbf{Primitive element:} A primitive element (generator) of $\mathbb{F}_5$ is $\gamma = 2$. To check this, we compute its powers modulo 5:
    \[
    2^1 = 2, \quad 2^2 = 4, \quad 2^3 = 8 \equiv 3, \quad 2^4 = 16 \equiv 1.
    \]
    The resulting set $\{2, 4, 3, 1\}$ contains all non-zero elements of $\mathbb{F}_5$. Therefore, $\gamma=2$ is a primitive element. In contrast, $\gamma=4$ is not, as it only generates the subset $\{4, 1\}$.
    \item \textbf{Polynomials:} Consider a polynomial $ P(x) = -7x^3 + 12x^2 - 4x + 10 \pmod 5 $ with $x \in \mathbb{F}_5$.  Under the modular map $ \pi $, this is equal to $ Q(x) = \pi ( P(x)) =  3x^3 + 2x^2 + x + 0 \pmod 5 $. 
\end{itemize}
\end{example}

\section{Max-LINSAT}
\label{sec:optimization}

In this section, we introduce the max-LINSAT problem --- the particular optimization problem that DQI is capable of solving.  

Max-LINSAT is defined as follows. Consider a vector of $ n $ variables $ \vec{x} =(x_1, x_2, \dots , x_n ) \in \mathbb{F}_p^n $.  The vector $ \vec{x} $ is then subject to $ m $ linear constraints of the form
\begin{align}
\label{maxlinsat}
   \vec{b}_i \cdot \vec{x} =  \sum^n_{j=1} B_{ij}x_j   \in T_i ,
\end{align}
where $ i \in [1,m] $. The aim is to find the vector $ \vec{x} $ satisfying as many as possible of these $m$ constraints.  Here, $ B_{ij} \in \mathbb{F}_p$ are elements of a $ m \times n $ rectangular matrix $ B $. 
The vector $ \Vec{b}_i $ is the $i$th row of the $B $ matrix. 
Each $ T_i =\{v_1, v_2, \dots, v_r \}  \subseteq  \mathbb{F}_p $ defines a target set which each give a constraint. The set  $T_i  $ is not necessarily a single element so that (\ref{maxlinsat}) may be satisfied in multiple ways. In this review, we take the total number of allowed values to be a constant $ |T_i | = r $.


The max-LINSAT problem can be reformulated in terms of an objective function in the following way.  For the $ i$th condition of (\ref{maxlinsat}), define a function with domain $ y \in \mathbb{F}_p  $  
\begin{align}
    \label{fidef}
f_i(y) = \begin{cases}
        1 \text{ for $y \in T_i $} \\
        -1\text{ for $y \notin T_i$}
    \end{cases} .
\end{align}
%
%
%
The objective function is then defined as 
\begin{align}
    f(\vec{x}) = \sum_{i=1}^m f_i (\vec{b}_i \cdot \vec{x}  ) = \text{\#SAT} - \text{\#UNSAT} .
    \label{objectivefunc}
\end{align}
Since $ f_i = \pm 1 $, this is the difference between the number of satisfied constraints and unsatisfied constraints. 
Then max-LINSAT can be formulated as the optimization problem 
\begin{align}
    \arg\max_{\vec{x} \in \mathbb{F}_p^n } f(\vec{x}) .
\end{align}
The difficulty of the optimization problem arises from the large number ($ p^n $) of candidate solutions $ \vec{x} $.


There are several special cases of max-LINSAT which we also consider.  The first is max-XORSAT, where the variables are in $\mathbb{F}_2 $, i.e. binary variables.  This limits the number of allowed values to $ r = 1 $, so that $ T_i = \{ v_i \} $.  The constraint equations in this case can be written
\begin{align}
\label{maxxorsat}
    \vec{b}_i \cdot \vec{x} =\sum^n_{j=1} B_{ij}x_j = v_i.
\end{align}
In the objective function formulation, we may explicitly write
\begin{align}
\label{maxxorsatfi}
f_i(y) = (-1)^{y + v_i} ,
\end{align}
where $ y \in \mathbb{F}_2 $ and addition is modulo 2.  
The total objective function to be maximized is then
\begin{align}
\label{objfunc}
f(\vec{x} ) = \sum_{i=1}^m (-1)^{\vec{b}_i \cdot \vec{x} + v_i} .
\end{align}

We also define the Optimal Polynomial Intersection (OPI) problem, which is another special case of max-LINSAT.  The OPI problem considers polynomials of degree at most $n-1$ in one variable $ y \in \mathbb{F}_p $ with coefficients $ x_j \in \mathbb{F}_p $ for $ j \in [0,n-1] $
\begin{align}
Q(y) = x_0 + x_1 y + \dots + x_{n-1} y^{n-1} .
\label{polyq}
\end{align}
The aim is to find the polynomial that maximizes the number of intersections with target sets $ T_i $ 
\begin{align}
\label{opimaximize}
\max_{Q(y)} |\{i  \in [1,m] : Q(y_i) \in T_i\}| ,
\end{align}
where $ y_i \in  \mathbb{F}_p $ with $ i \in [1,m] $ are $m$ evaluation points of $ y $ (i.e. $i$ labels the $m$ possible values of $y$), with typically $ m = p-1 $. The number of polynomial coefficients $n$ is taken such that it is less than or equal to $ m $, i.e. $ 0 \le n \le m < p $, making this constraint system overdetermined.

We may formulate OPI in max-LINSAT form (\ref{maxlinsat}) in the following way.  Define the $ B $ matrix with elements $ B_{ij} = y_i^j $ in  Vandermonde form
\begin{align}
    B = 
\begin{pmatrix}
  1 & y_1 & y_1^2 &\dots & y_1^{n-1} \\
  1 & y_2 & y_2^2 &\dots & y_2^{n-1} \\
   \vdots & \vdots & &  & \vdots \\
   1   & y_m &  y_m^2 &\dots & y_m^{n-1} \\ 
\end{pmatrix}   ,
\label{Bvoldemortmat}
\end{align}
where $ y_i \in  \mathbb{F}_p $ as before.  The dot product of the $i$th row of the matrix with the vector $ \vec{x} $ recovers the polynomial (\ref{polyq}) evaluated at $ y_i $, i.e. $ \vec{b}_i \cdot \vec{x} = Q(y_i) $.  Maximization of the objective function (\ref{objectivefunc}) then recovers the best polynomial fit to the target data $ T_i $.  Example \ref{example:OPI_instance} shows an instance of the OPI problem.

The max-LINSAT, max-XORSAT, and OPI have a worst-case complexity that is NP-hard for exact optimization. Their decision problem versions are NP-complete.  Intuitively, OPI is NP-hard because selecting coefficients to maximize the intersections is a constrained combinatorial problem.





\begin{example}{OPI instance}{OPI_instance}
    Here we give an explicit example of the OPI problem. We  assume the following parameters.  We work with variables in $\mathbb{F}_7$, i.e. $p$ = 7.  The largest degree of the polynomial $Q(y) $ is 2, such that $ n = 3 $.  The number of constraints is $ m = 6 $, with $T_1 = \{0, 1\}, T_2 = \{3, 6\}, T_3 = \{2, 5\}, T_4 = \{3, 6\}, T_5 = \{4, 5\}, T_6 = \{1, 5\}$.  Each constraint set $T_i $ is of size $ r =2 $. 

    The Vandermonde matrix $B \in \mathbb{F}_p^{m\times n} = \mathbb{F}_7^{6 \times 3}$ is equal to
\begin{align}
\renewcommand{\arraystretch}{0.5}
    B = 
    \begin{pmatrix}
    1 & 1 & 1\\
    1 & 2 & 4\\
    1 & 3 & 2\\
    1 & 4 & 2\\
    1 & 5 & 4\\
    1 & 6 & 1
    \end{pmatrix} ,
\end{align}
%
where we took $ y_i = i \in  [1,6 ] $, $j  \in [0,2] $, and $ B_{ij} = y_i^j \text{ mod 7}$. 

The objective function (\ref{objfunc}) in our case is
\begin{align}
  f(\Vec{x})= \sum_{i=1}^{6} f_i( \Vec{b}_i \cdot \Vec{x}) = \sum_{i=1}^{6} f_i( Q(y_i)), 
  \label{fopi}
\end{align}
where $\Vec{x} = (x_0,  x_1, \dots,  x_{n-1} ) $ is a vector containing the polynomial coefficients $x_i $, and $ \Vec{b}_i $ is the $i$th row of the $B $ matrix.  
The functions $f_i $ are defined according to (\ref{fidef}).  For example, $ f_1(y) = 1  $ for $ y \in  T_1 = \{0, 1\} $, and $ f_1(y) = - 1  $ for $ y  \in  \mathbb{F}_7 \backslash T_1 = \{2, 3, 4, 5, 6\} $.

The key purpose of matrix $B$ here is linking polynomial coefficients $\Vec{x}$ to the evaluations $Q(y)$ of inputs $y \in \mathbb{F}_7$. For example, for the polynomial coefficients 
$\Vec{x} = (4,1,2) $, 
the polynomial is 
$Q(y) = 4+ y + 2 y^2$.
Given the $i$th input $y_i$, evaluating the polynomial $ Q(y_i) $ corresponds to taking the dot product with the $i$th row of the $ B $ matrix.  For $ y = 3 $, taking the dot product with the 3rd row of $ B $, we may evaluate
\begin{align}
        Q(3) &= 4 B_{30} + B_{31} +2 B_{32}  = 11 \equiv 4 \pmod{7} \notin T_3 \implies\ f_3 = -1 .
\end{align}
Evaluating the objective function for all $ i \in [1,6] $, we have $ f_i = \{ 1 , -1 , -1 , -1 , -1, 1 \} $. Then we may evaluate (\ref{fopi}) as  $f(\vec{x}) = 2-4 = -2$.

For this example, no perfect polynomial ($f(\vec{x} ) = m = 6$) exists that is a degree at most 2 polynomial. Hence, the optimal polynomial would be 
$ Q^* (y) = 1 + 5y + 5y^2$ 
and $f(Q^*) = 5 -1 = 4$.
    
    \begin{figure}[H]
        \centering
        \includegraphics[width=0.4\linewidth, height=0.4\linewidth]{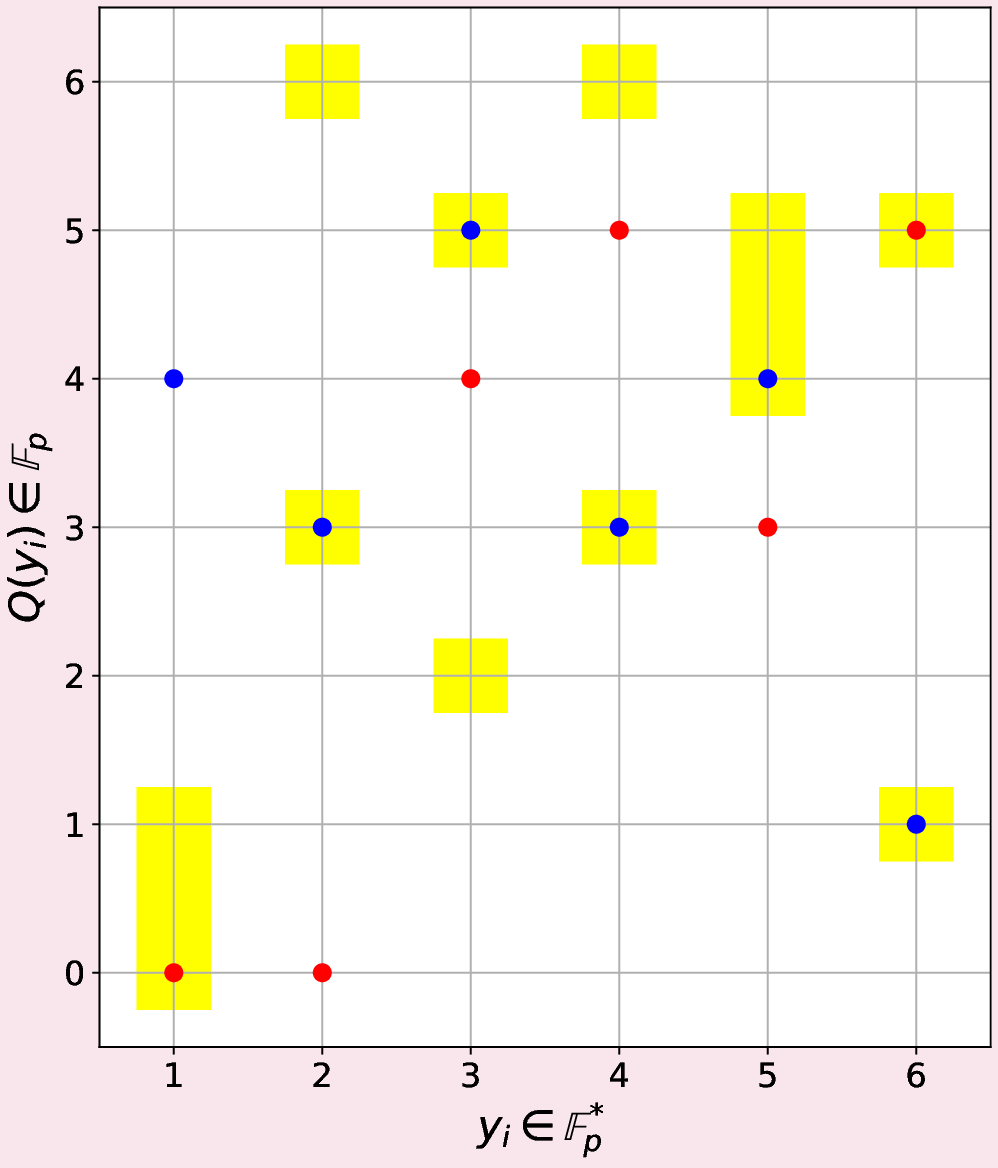}
        \caption{\textbf{Visualization of the OPI problem.} We show the polynomial $Q(y) = 4 + y + 2y^2 $ (red) and the optimal polynomial $Q^*(y) = 1 + 5y + 5y^2 $ (blue). Yellow boxes indicate the target sets $ T_i $. }
        \label{polynomial_visualization}
    \end{figure}

\end{example}

\section{Coding theory}
\label{sec:coding}

In this section, we give a brief description of coding theory.  Coding theory can be considered the mathematical framework to enable sending (classical) information through a noisy channel, i.e. error detection and correction.  By encoding messages in terms of codewords that include redundancy, the objective is to detect these errors and, when possible, recover the original message by decoding the received message.  This is done by designing the code so that received words can be reliably decoded back into the intended codeword.  For a more detailed description we refer the reader to the coding theory preliminaries given in Ref. \cite{c_t_soft}.


Suppose we have a message $ \vec{m}  $ that is a vector of length $ k $, and each component $ m_i \in \mathbb{F}_p $  for $ i \in [0,k-1] $ is a symbol from the alphabet $ \mathbb{F}_p  $ 
\begin{align}
\vec{m} = ( m_0, m_1, \dots, m_{k-1} ) .
\end{align}
One of the central tasks in coding theory is to encode this message as a codeword.  There are a total of $ p^k $ possible messages, hence the set of all codewords $ \vec{c}_i $ which together form the \textit{code}
\begin{align}
    C=\{ \vec{c}_1, \vec{c}_2, \dots , \vec{c}_{p^k} \} .
\end{align}
These codewords are transmitted through a channel, and later decoded so that the original message $ \vec{m} $ is recovered.  As we discuss further below, the codewords have properties such that the decoding can proceed even in the presence of noise.  

We will consider \textit{block codes}  where codewords are vectors of the same length $ n $ and each component $ c_i $ is a symbol from the  alphabet $  \mathbb{F}_p $
\begin{align}
\vec{c} = ( c_0, c_1, \dots, c_{n-1} ) .
\end{align}
Here, $ n > k $ such that the codewords are longer than the messages themselves.  This redundancy is what allows for error detection and correction. 

\textit{Linear codes} are codes where any linear combination of codewords is also a codeword, i.e. $ a \vec{c} + b \vec{d} \in C $, for $ a,b  \in \mathbb{F}_p $ and $ \vec{c} , \vec{d} \in C $.  A \textit{dual code} $C^{\perp}$ of a linear code  $C$ is the set of all vectors that are orthogonal to every codeword in $C$
\begin{align}
C^{\perp} = \{ \vec{c}^{\perp} \in \mathbb{F}_p^n : \langle  \vec{c}^{\perp},\vec{c} \rangle = 0 \hspace{2mm} \forall \vec{c}  \in C  \} .   
\end{align}
Here, $ \langle  \vec{b}, \vec{c} \rangle $ is an inner product, which for our purposes we take to be the dot product 
$ \langle  \vec{b}, \vec{c} \rangle  = \vec{b} \cdot \vec{c} \pmod p $.  The size of the dual code is $ | C^{\perp} | = p^{n-k} $.  Linear block codes are typically denoted by $ [n,k] $, where $n$ represents the codeword length and $k$ represents the dimension of the code (the number of symbols).  The \textit{minimum distance} $d$ is the smallest Hamming distance between any two distinct codewords in the code $C$. 

To encode the message $ \vec{m} $, the generator matrix $  G \in \mathbb{F}^{k\times n}_p $ is used to encode the message into a codeword according to
\begin{align}
\label{encodingstep}
    \vec{c} = \vec{m} G
\end{align}
The generator matrix has dimensions $ k \times n $, such that the output codeword $ \vec{c} $  has a length $ n $. The codeword is then transmitted through a noisy channel.  During this process, the received codeword may become corrupted according to
\begin{align}
\label{corruptedcodeword}
    \vec{y} = \vec{c} + \vec{e}  ,
\end{align}
where $ \vec{e} $ is an error vector.  In a \textit{syndrome decoder}, to determine the error one makes use of the parity-check matrix $H \in \mathbb{F}^{(n-k)\times n}_p$ which has the property $ G H^T = 0 $, so that for any codeword
\begin{align}
\label{paritycheckcond}
    \vec{c} H^T = 0 .
\end{align}
This means that when applied to our corrupted codeword we have
\begin{align}
\label{syndrome}
    \vec{s} = \vec{y} H^T = \vec{e} H^T ,
\end{align}
which is the syndrome.  Solving for $ \vec{e} $ (which is in general a non-trivial problem), we may recover the original codeword by $ \vec{c} = \vec{y} - \vec{e} $.  Finally, the original message is obtained by inverting (\ref{encodingstep}), performed by decoding.   As we discuss in more detail below, the practical use of a particular code depends on whether such a decoding process can be efficiently be performed.  



We now discuss some specific examples of codes that are relevant for the DQI algorithm.  The \textit{Reed-Solomon} (RS) code is constructed by forming a polynomial of degree at most $ k -1 $ defined over  $ x \in \mathbb{F}_p $, where the coefficients correspond to the message $ \vec{m} $
\begin{align}
    P(x) = m_0 + m_1 x + \dots + m_{k-1} x^{k-1}  .
\end{align}
Then a codeword is created by picking $ n $ distinct points $ x_i \in \mathbb{F}_p $ on the polynomial
\begin{align}
\label{rsencoding}
    \vec{c} = ( P(x_1), P(x_2), \dots, P(x_n) ) .
\end{align}
Choosing $ n > k $ points creates a natural redundancy because a polynomial of degree $ k - 1 $ is determined by $ k $ distinct points.  Without errors, the $n-k $ points provide redundant information about the polynomial.  When errors are present, one can take advantage of the redundancy to detect and correct errors.  The above procedure can be expressed in coding-theoretic form as follows.  For the above encoding step the associated generator matrix takes a Vandermonde form
\begin{align}
\renewcommand{\arraystretch}{0.5}
    G = 
\begin{pmatrix}
  1 & 1 & \dots &  1\\
  x_1 & x_2 & \dots & x_n \\
   x_1^2 & x_2^2 & \dots & x_n^2 \\
   \vdots & \vdots &   & \vdots \\
     x_1^{k-1} & x_2^{k-1} & \dots & x_n^{k-1} \\ 
\end{pmatrix}   .
\label{voldemortmat}
\end{align}
Applying (\ref{encodingstep}) produces the same encoding as (\ref{rsencoding}).  In Example \ref{example:reedsolomon} we show a concrete example of an RS code.

\begin{example}{Reed-Solomon code RS[7, 3] over $\mathbb{F}_7$}{reedsolomon}
We consider an  RS[7, 3] code, such that the codeword length is $ n =7 $ and $k =3 $ is the message length.  Each component of the codeword $ \vec{c} $ and the message $ \vec{m} $ is a symbol from the alphabet $ \mathbb{F}_7 $, which consists of 7 symbols.  The total number of messages is therefore $ p^k = 7^3 = 343 $, and there is a codeword associated with each one, hence $ | C | = 343 $.  

The RS polynomial that we use in this case is
\begin{align}
\label{examplerspoly}
    P(x) = m_0 + m_1 x + m_2 x^2 .
\end{align}
This is used to generate the codeword associated with the message by choosing $ n=7 $ points on the polynomial.  Here we take this to be $ x_i = i \pmod 7 $ where $ i \in [1,7] $, so that
\begin{align}
    \vec{c} = (P(1), P(2),P(3),P(4),P(5),P(6),P(7) ) .
\end{align}
Three points fully specify the polynomial, and from this the message $ \vec{m} $ can be deduced in the error-free case.  The additional points then provide redundancy that enables error detection.  The minimum distance for this code is $ d = n - k + 1 = 5 $.  

The encoding process may equally be evaluated by using (\ref{encodingstep}).  The generator matrix in our case is
\begin{align}
\renewcommand{\arraystretch}{0.5}
G =    \begin{pmatrix}
    1 & 1 & 1 & 1 & 1 & 1 & 1\\
    1 & 2 & 3 & 4 & 5 & 6 & 0 \\
    1 & 4 & 2 & 2 & 4 & 1 & 0
    \end{pmatrix} .
\end{align}
For example, evaluating the message polynomial with coefficients $\vec{m} = (3, 5, 1)$ produces the codeword $\vec{c} = (2,3,6,4,4,6,3)$. 
\end{example}

The \textit{Reed-Muller} (RM) code is also based upon polynomials, but uses multivariate Boolean polynomials in variables $ x_i \in \mathbb{F}_2 $ with $ i \in [1,m] $ up to degree $ r $
\begin{align}
P(x_1,x_2,\dots, x_m) & = \sum_{j=0}^{k_r-1} m_j x_1^{b^{(1)}_j} x_2^{b^{(2)}_j} \dots x_m^{b^{(m)}_j} \nonumber \\
& = m_0 + m_1 x_1 + m_2 x_2 + \dots + m_m x_m + m_{m+1} x_1 x_2 + m_{m+2} x_1 x_3  + \dots
\label{rmpoly}
\end{align}
Here, $ b^{(i)}_j \in \{0,1 \} $ returns the $ i$th digit of an $m$-digit binary number $b_j $ ordered by the number of 1's.  For example, for $ m=3 $, $ b_0 = 000, b_1 = 100, b_2 = 010, b_3 = 001, b_4=110, b_5 = 101, b_6 =011, b_7 =111 $. The number of terms in the sum in (\ref{rmpoly}) is 
\begin{align}
    k_r = \sum_{l=0}^r \binom{m}{l} ,
\end{align}
such that $ \sum_{i=1}^m b^{(i)}_j \le r $.  The coefficients of the polynomial correspond to the message  $ \vec{m} = (m_0, m_1, \dots, m_{k_r - 1} ) $.  The codewords are formed by evaluating over all $ n = 2^m $ values of $ \vec{x} = (x_1, x_2, \dots, x_m ) $ 
\begin{align}
\vec{c} = ( P(0,\dots,0,0), P(0,\dots,0,1), P(0,\dots,1,0), \dots , P(1,\dots,1,1)) .
\end{align}
RM codes are typically denoted in terms of the polynomial degree and number of variables and written as $ \text{RM}(r,m)$.  An explicit example is given in Example \ref{example:example2}.

\begin{example}{Reed-Muller code RM(1,3) over $\mathbb{F}_2$}{example2}
We consider a RM(1,3) code, corresponding to multivariate polynomials up to degree $ r = 1 $ with $ m = 3 $ variables. We work with binary variables in $ \mathbb{F}_2  $, so that $ p =2 $.   This corresponds to polynomials of the form
\begin{align}
P(x_1, x_2, x_3) = m_0 + m_1 x_1 + m_2 x_2 + m_3 x_3 .
\end{align}
The coefficients of the polynomial form the message vector $ \vec{m} = (m_0, m_1, m_2, m_3 ) $, so that the message length is $ k_r = 4 $. The codeword is formed by evaluating over all values of $ \vec{x} = (x_1, x_2, x_3) $:
\begin{align}
\vec{c} = (P(0,0,0), P(0,0,1), P(0,1,0), P(0,1,1), P(1,0,0), P(1,0,1), P(1,1,0), P(1,1,1)) .
\end{align}
The codeword length is therefore $ n = 2^m = 8 $.  Hence the RM(1,3) code corresponds to a $ [n,k] = [8,4]$ binary linear code. The total number of messages (and hence codewords) is $ p^{k_r} = 16 $. The minimum distance is $ d = 2^{m-r} = 4 $.  



As before, we may construct a generator matrix to perform the encoding process.  In our case the columns are the values $ ( 1, x_1, x_2, x_3)^T $ for all values of $ \vec{x} $:
\begin{align}
G =\renewcommand{\arraystretch}{0.5}
    \begin{pmatrix}
    1 & 1 & 1 & 1 & 1 & 1 & 1 & 1\\
    0 & 0 & 0 & 0 & 1 & 1 & 1 & 1\\
    0 & 0 & 1 & 1 & 0 & 0 & 1 & 1\\
    0 & 1 & 0 & 1 & 0 & 1 & 0 & 1
    \end{pmatrix} .
\end{align}


For example, for $\Vec{m} = (1, 0, 1, 1) $, using (\ref{encodingstep}), 
%
we obtain the codeword $ \vec{c} = (1, 0, 0, 1, 1, 0, 0, 1) $.  
\end{example}

As mentioned above, the central task of coding theory is, through the use of decoding algorithms, to recover the original, intended message from a received word that may contain errors.  The RS and RM codes are interesting because of their algebraic structure and error-correcting capability: there exist efficient algorithms which can correct up to $t = \lfloor \frac{d-1}{2} \rfloor $ symbol errors. In fact, the decoding problem has a similar structure to max-LINSAT that was described in Sec. \ref{sec:optimization}. Two of the groups of decoding problems are Bounded Distance Decoding and Approximate Nearest Codeword Decoding which are NP-complete and NP-hard problems, respectively. However, in practice, for very structured codes, there exist efficient decoding algorithms. This is the crucial fact that eventually allows DQI to be efficiently implementable. The Berlekamp-Massey decoding algorithm \cite{Berlekamp2015} is a key polynomial-time decoder of Reed–Solomon codes.  In Example \ref{example:example3}, we give a simplified example of the Approximate Nearest Codeword Decoding algorithm which is applicable for the RM code.


\begin{example}{Decoding algorithm}{example3}
\begin{figure}[H]
    \centering
    \includegraphics[width=0.5\linewidth, height=0.43\linewidth]{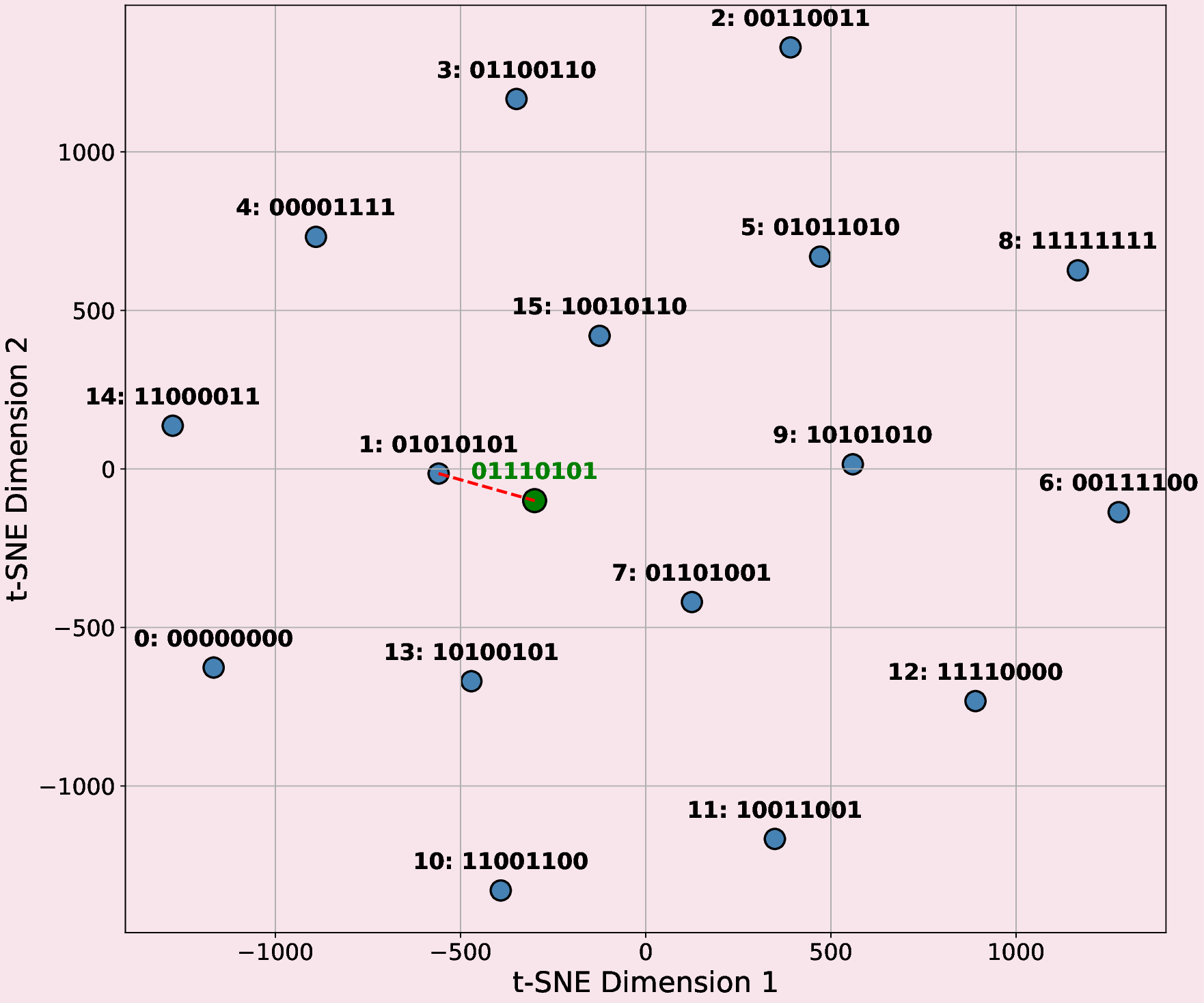}
    \caption{\textbf{Approximate Nearest Codeword Decoding of a RM(1,3) code. } The message $\Vec{m} = 01110101$ (shown in green) is decoded to the closest of the available 16 codewords (blue points) from the RM(1,3) code in Example \ref{example:example2}. 
    The 2D visualization of the codewords was done with the help of t-SNE  embedding.
    \label{fig:decodingrmcode}}
\end{figure}
We consider the decoding process in a  RM(1,3) code.  In Fig. \ref{fig:decodingrmcode}, the codewords are represented as points in the ambient space.  A message experiences some noise which causes it to deviate from the codewords.  Using the Hamming distance criterion  $\arg \min_{\Vec{x}\in C} |\Vec{m}-\Vec{x}|$, the corrupted message is decoded to the codeword $\Vec{x}_1 = 01010101$.
\end{example}


\section{Decoded Quantum Interferometry}
\label{DQI_explanation}

This section introduces DQI, the central quantum algorithm that is discussed in this review. We first make the distinction between exact and approximate optimization to clarify the nature of the task that DQI achieves.  We then discuss the DQI target state, i.e. the quantum state that the DQI algorithm aims to prepare, which shows the way in which the solution of max-LINSAT is approximated.  Then we give a step-by-step breakdown of how to prepare this state for the simpler case of max-XORSAT. Finally, we describe the changes required for the more general case of max-LINSAT.


\subsection{Exact vs. Approximate Optimization\\}

As discussed in Sec. \ref{sec:optimization}, general max-LINSAT and its special cases max-XORSAT and OPI are NP-hard problems in the worst-case for exact optimization. 
For such problems there are no known polynomial-time algorithms  --- classical or quantum --- which output the exact optimal solution.  There are however algorithms which can give {\it approximate} solutions which are not necessarily optimal, but give high-quality solutions.  To compare such approximate algorithms, the performance measure used to evaluate algorithms is the \textit{expected constraint satisfaction}  $\langle s\rangle$. As the name suggests, this is the average number of constraints that are satisfied. This allows for a way to compare different algorithms and compare their performance. 

When comparing different algorithms, one typically must also take into account of the resources required to attain a particular $\langle s\rangle$.  Typically approximate algorithms possess an accuracy parameter, which allows one to increase $\langle s\rangle$ in exchange for a longer run time.  For example, in annealing algorithms (quantum or classical) a better quality solution can be  attained by a slower annealing schedule, which increases the run time of the algorithm.  Thus in any comparison, one must either fix the resources (i.e. time) to attain the solution and compare the $\langle s\rangle$, or alternatively compare the resources needed to attain the same $\langle s\rangle$.  

\begin{figure}[t]
    \centering
    \includegraphics[width=0.5\textwidth]{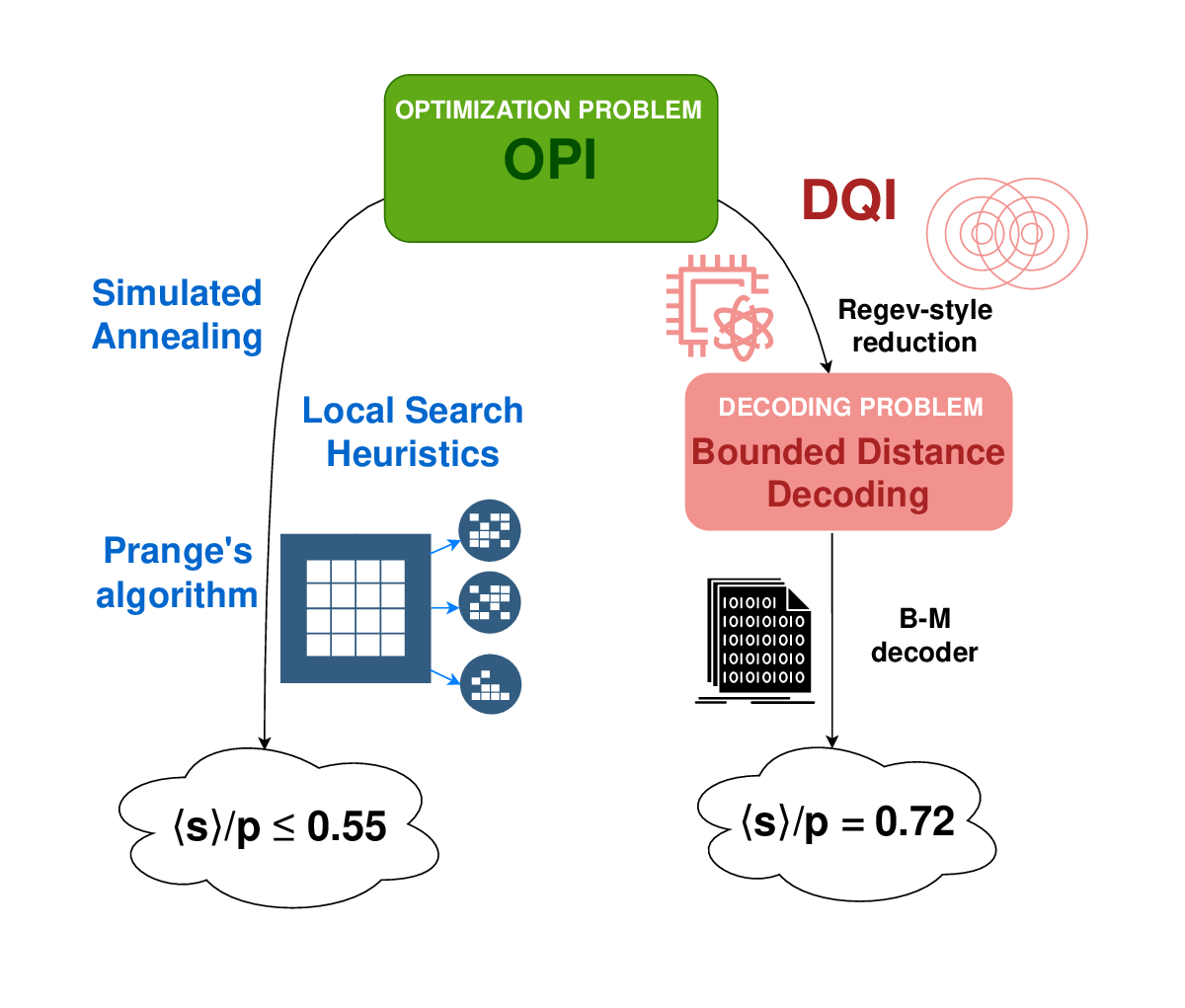}
    \caption{ \textbf{Approaches for tackling the OPI optimization problem.}   Shown are the best-known classical methods (left) and the quantum DQI  approach (right). Using DQI, a higher constraint satisfaction rate can be attained using quantum interference and classical decoding methods.}
    \label{fig:sexp}
\end{figure}

In the case of DQI, the aim is to approximate the solution of the max-LINSAT problem.  The best performance that was observed was for the OPI problem, a special case of max-LINSAT as discussed in Sec. \ref{sec:optimization}.  The best known polynomial-time classical algorithm in this case is Prange's algorithm.  What was found in Ref. \cite{jordan2024} is that DQI gives a superior expected constraint satisfaction rate of $ \frac{\langle s \rangle}{p} = 0.72 $ in comparison to Prange's algorithm benchmark of 0.55 (see Fig. \ref{fig:sexp}).  What is claimed in Ref. \cite{jordan2024} is that for the classical algorithm to match the same $\langle s\rangle$ as DQI, a superpolynomial amount of time would be required.  This is the nature of the quantum speedup in DQI.

\subsection{DQI target state\\}


As discussed in Sec. \ref{sec:optimization}, max-LINSAT can be written as a maximization of the objective function $ f(\vec{x}) $, given by Eq. (\ref{objectivefunc}).  
One way to use a quantum computer to solve max-LINSAT would be to prepare the state 
\begin{align}
  \propto  \sum_{\Vec{x} \in \mathbb{F}_p^n} (f (\Vec{x})  +m) \ket{\vec{x}} .
\end{align}
Here, the sum runs over all $ \Vec{x} $ that give candidate solutions.  The $ +m$ term above guarantees that the amplitude is a non-negative quantity since $ f \in [-m,m] $.  Then a measurement in the $ \ket{\vec{x}} $ basis yields the outcome $ \vec{x} $ with probability $ \propto (f(\Vec{x})+m )^2$.  
We can however do better than this by applying a power to the amplitude by preparing $  \propto \sum_{\Vec{x} \in \mathbb{F}_p^n} (f(\Vec{x})  +m)^\ell \ket{\vec{x}} $.  Then since $(f(x)+m)^{2\ell} \ge (f(x)+m)^2$ for integers, we will obtain the best solutions with a higher relative probability compared to lower scoring solutions.


More generally, we can consider a degree-$\ell$ polynomial called the amplifying function (also called the retroaction polynomial)
\begin{align}
\label{retroactionpoly}
    P(f( \vec{x}) ) = \sum_{k=0}^\ell \alpha_k f^k ( \vec{x}) .
\end{align}
Then we may define the DQI state
\begin{align}
\label{dqistate}
    \ket{\text{DQI}} = \ket{P(f)} := \displaystyle \sum_{\Vec{x} \in \mathbb{F}_p^n} P(f ( \vec{x}) )\ket{\vec{x}} , 
\end{align}
which on measurement in the $ \ket{\vec{x}}$ basis results in an enhancement of higher-scoring solutions for suitable choices of $ \alpha_k $.  Here, we include any normalization factors  in $ \alpha_k $  such that the DQI state is normalized. 

The aim of the DQI algorithm is to prepare the DQI state (\ref{dqistate}). To this end, we write the amplitude of the DQI state in an alternative way.  Using the fact that the objective function is a simple sum of the individual constraint functions $ f = \sum_{i=1}^m f_i $, we may rewrite (\ref{retroactionpoly}) 
\begin{align}
\label{symmpolydecomp}
      P(f ) = \sum_{k=0}^\ell u_k P^{(k)} ( f_1,    f_2, \dots,  f_m )  ,
\end{align}
where we have omitted the $  \vec{x} $ dependence on the functions for brevity and $ P^{(k)} ( f_1,    f_2, \dots,  f_m ) $ are symmetric polynomials of degree $ k $.  A degree-$k$ elementary symmetric polynomial $P^{(k)}(f_1, \dots, f_m)$ is an unweighted sum of all $\binom{m}{k}$ products of $k$ distinct factors from $f_1, \dots, f_m$. The coefficients $ u_k $ are new expansion coefficients to be determined from $ \alpha_k $. 
For example, Table \ref{tab:sympoly} shows the symmetric polynomials for $ m = 4 $. 

\begin{table}[t]
\begin{center}
\small
\begin{tabular}{|c|c|c|}
\hline
$k$ & Symmetric polynomial $P^{(k)}$ & Dicke state $ |D^m_k \rangle $ \\
\hline
0 & $1$ & $ | 0000 \rangle $ \\
1 & $ f_1 + f_2 + f_3 + f_4  $ & $ \frac{1}{2} (| 1000 \rangle + | 0100 \rangle  + | 0010 \rangle  + | 0001 \rangle ) $  \\
2 & $ f_1 f_2 + f_1f_3 + f_1f_4 + f_2f_3 + f_2f_4 + f_3f_4$ & $ \frac{1}{\sqrt{6}} ( | 1100 \rangle + | 1010 \rangle  + | 1001 \rangle  + | 0110 \rangle + | 0101 \rangle + | 0011 \rangle ) $  \\
3 & $ f_1f_2f_3 + f_1f_2f_4 + f_1f_3f_4 + f_2f_3f_4$ & $ \frac{1}{2} ( | 1110 \rangle + | 1101 \rangle + | 1011 \rangle + | 0111 \rangle )$  \\
4 & $ f_1f_2f_3f_4$ & $ | 1111 \rangle $  \\
\hline
\end{tabular}
\captionof{table}{Elementary symmetric polynomials for $m=4$. Also shown are the associated Dicke states, which are symmetric states with exactly $ k $ excitations. \label{tab:sympoly}}
\end{center}
\end{table}

The property $f_i^2 = 1$ is key to reducing higher powers of $f_i$ to linear terms, e.g. $f_i^3 = f_i, f_i^4 = 1$ and so on,  due to the fact that $f_i =\pm 1 $. For example, for $ m = 2 $ and $ \ell = 2 $, we may write
\begin{align}
 P(f) & = \alpha_0 + \alpha_1 ( f_1 + f_2 ) + \alpha_2 ( f_1 + f_2 )^2 \nonumber \\
 & = \alpha_0 + 2 \alpha_2 + \alpha_1 ( f_1 + f_2 ) + 2 \alpha_2 f_1 f_2  .
\end{align}
The $ u_k $ coefficients in this case are $ u_0 =  \alpha_0 + 2 \alpha_2 $, $ u_1 = \alpha_1 $, $ u_2 = 2 \alpha_2 $.  

Each $f_i$ represents the satisfaction of a constraint (taking the value $+1$ when satisfied and $-1$ otherwise). 
The  product $f_{i_1} f_{i_2} \dots f_{i_k}$ encodes what we refer to as the joint satisfaction correlations of $k$ specific constraints.
This product taking a value $+1$ does not imply all $k$ constraints are satisfied --- we only gain information about the parity of that combination of constraints (whether an odd or an even number of them are satisfied). Constructive interference between states encoding the joint satisfaction correlations of constraints 
is what gives rise to the  amplification of good solutions in DQI.  Further aspects of symmetric polynomials and other details can be found in Appendix \ref{Elementary_symmetric_polynomials}.

Substituting (\ref{symmpolydecomp}) into (\ref{dqistate}) we then obtain the form of the DQI state that is prepared in the  DQI algorithm
\begin{align}
\label{dqistate2}
    \ket{\text{DQI}} = \ket{P(f)} = \displaystyle \sum_{\Vec{x} \in \mathbb{F}_p^n} \sum_{k=0}^\ell u_k P^{(k)} ( f_1,    f_2, \dots,  f_m )   \ket{\vec{x}} . 
\end{align}

\subsection{DQI for max-XORSAT\\}
\label{sec:dqimaxxorsat}

Next, we go through the seven steps of the DQI algorithm. 
We will first consider solving the max-XORSAT problem, which is a special case of max-LINSAT.  In later sections we discuss how this is generalized to max-LINSAT.  

First, let us write the DQI state for the case of max-XORSAT starting from the expression (\ref{dqistate2}).  First, note that the symmetric polynomial can be written as
\begin{align}
P^{(k)} ( f_1,    f_2, \dots,  f_m )  = \sum_{| \vec{y} | = k } (f_1)^{y_1} (f_2)^{y_2} \dots (f_m)^{y_m} ,
\end{align}
where $ \vec{y} $ is a binary vector of length $ m $.  In the sum, we restrict to vectors with a Hamming weight that is equal to $ k $, i.e. those with $ k $ elements taking the value 1 and the remaining are 0.  Substituting in the expression (\ref{maxxorsatfi}), we obtain the explicit expression
\begin{align}
P^{(k)} ( f_1,    f_2, \dots,  f_m )  = \sum_{| \vec{y} | = k } (-1)^{ \vec{v} \cdot \vec{y} + (B^T \vec{y} ) \cdot \vec{x} } .
\end{align}
Substituting this into (\ref{dqistate2}), we have
\begin{align}
   \ket{P(f)} = \sum_{k=0}^\ell u_k  \sum_{| \vec{y} | = k } \sum_{\Vec{x}}  (-1)^{ \vec{v} \cdot \vec{y} + (B^T \vec{y} ) \cdot \vec{x} }    \ket{\vec{x}} , 
\end{align}
where $ \vec{x} $ is a binary vector of length $ n $ in this case.  In order that this is a normalized state the coefficients $ u_k $ must satisfy
\begin{align}
  2^n  \sum_{k=0}^\ell |u_k|^2 \binom{m}{k} = 1  . 
\end{align}
It is therefore convenient to define the coefficients
\begin{align}
    w_k = u_k \sqrt{2^n \binom{m}{k}}
\end{align}
such that $ \sum_k |w_k |^2 = 1 $ instead.  We then obtain the target DQI state for max-XORSAT
\begin{align}
\label{dqistatexorsat}
    \ket{\text{DQI}} = \ket{P(f)} = \frac{1}{\sqrt{2^n}}  \sum_{k=0}^\ell \frac{w_k}{\sqrt{\binom{m}{k}}}  \sum_{| \vec{y} | = k } \sum_{\Vec{x}}  (-1)^{ \vec{v} \cdot \vec{y} + (B^T \vec{y} ) \cdot \vec{x} }    \ket{\vec{x}} . 
\end{align}

\begin{figure}[t]
    \centering
    \begin{tikzpicture}
        \node at (-3,-3) {
            \hspace{-5cm}
            \centering
            \fontsize{9pt}{10pt}\selectfont
            \Qcircuit @C=1.0em @R=1.0em @! R {
                \lstick{\ket{0}} & \multigate{3}{\shortstack{$U_1$: \\ Preparing \\ weights $w_i$}} & \multigate{6}{\shortstack{$U_2$: \\ Preparing \\ Dicke states}} & \multigate{6}{\shortstack{$U_3$: \\ Uncomputing \\ weight \\ register}} & \qw \\ 
                \lstick{\ket{0}} & \ghost{\shortstack{$U_1$: \\ Preparing \\ weights $w_i$}} & \ghost{\shortstack{$U_2$: \\ Preparing \\ Dicke states}} & \ghost{\shortstack{$U_3$: \\ Uncomputing \\ weight \\ register}} &\qw \\ 
                \lstick{\vdots} \\
                \lstick{\ket{0}} & \ghost{\shortstack{$U_1$: \\ Preparing \\ weights $w_i$}} & \ghost{\shortstack{$U_2$: \\ Preparing \\ Dicke states}} & \ghost{\shortstack{$U_3$: \\ Uncomputing \\ weight \\ register}} & \qw \\
                \lstick{\ket{0}} & \qw & \ghost{\shortstack{$U_2$ \\ Preparing \\ Dicke states}} & \ghost{\shortstack{$U_3$: \\ Uncomputing \\ weight \\ register}} & \multigate{4}{\shortstack{$U_4$: \\ Encoding \\ target sets}} & \multigate{8}{\shortstack{$U_5$: \\ Matrix \\ multiplication}} & \multigate{8}{\shortstack{$U_6$: \\Uncomputing \\ error \\ register}} & \qw  & \qw\\
                \lstick{\vdots} & \vdots \\
                \lstick{\ket{0}} & \qw & \ghost{\shortstack{$U_2$ \\ Preparing \\ Dicke states}} & \ghost{\shortstack{$U_3$: \\ Uncomputing \\ weight \\ register}} & \ghost{\shortstack{$U_4$: \\ Encoding \\ target sets}} & \ghost{\shortstack{$U_5$:\\ Matrix \\ multiplication}} & \ghost{\shortstack{$U_6$: \\Uncomputing }} & \qw & \qw \\
                \lstick{\ket{0}} & \qw & \qw & \qw & \ghost{\shortstack{$U_4$: \\ Encoding \\ target sets}} & \ghost{\shortstack{$U_5$: \\ Matrix \\ multiplication}} & \ghost{\shortstack{$U_6$: \\Uncomputing}} & \qw & \qw\\
                \lstick{\ket{0}} & \qw & \qw & \qw & \ghost{\shortstack{$U_4$: \\ Encoding \\ target sets}} & \ghost{\shortstack{Matrix \\ multiplication}} & \ghost{\shortstack{Uncomputing}} & \qw & \qw \\
                \lstick{\ket{0}} & \qw & \qw & \qw & \qw & \ghost{\shortstack{Matrix \\ multiplication}} & \ghost{\shortstack{Uncomputing}} & \qw & \multigate{3}{\shortstack{$U_7$: \\ QFT}} & \meter \\
                \lstick{\ket{0}} & \qw & \qw & \qw & \qw & \ghost{\shortstack{Matrix \\ multiplication}} & \ghost{\shortstack{Uncomputing}} & \qw & \ghost{\shortstack{$U_7$: \\ QFT}} & \meter \\
                \lstick{\vdots} & \vdots & \vdots & \vdots & \vdots & \vspace{3cm} & \vspace{3cm} & \vspace{3cm} & \vspace{1cm} & \vdots \\
                \lstick{\ket{0}} & \qw & \qw & \qw & \qw & \ghost{\shortstack{Matrix \\ multiplication}} & \ghost{\shortstack{Uncomputing}} & \qw & \ghost{\shortstack{$U_7$: \\ QFT}} & \meter \\
            }
        };
        
        \draw[decorate, decoration={brace, mirror, amplitude=5pt}, thick] 
            (-14, 1.75) -- (-14, -0.95) node[midway, left=8pt] {\textsf{W}};
        \draw[decorate, decoration={brace, mirror, amplitude=5pt}, thick] 
            (-14, -1.2) -- (-14, -4.75) node[midway, left=8pt] {\textsf{E}};
        \draw[decorate, decoration={brace, mirror, amplitude=5pt}, thick] 
            (-14, -5) -- (-14, -7.75) node[midway, left=8pt] {\textsf{S}};

        \draw[dashed, blue, thick] (-11.135, 2) -- (-11.135, -7.75) node[below, black]{};
        \draw[dashed, blue, thick] (-8.95, 2) -- (-8.95, -7.75) node[below, black]{};
        \draw[dashed, blue, thick] (-6.575, 2) -- (-6.575, -7.75) node[below, black]{};
        \draw[dashed, blue, thick] (-4.675, -0.9) -- (-4.675, -7.75) node[below, black]{};
        \draw[dashed, blue, thick] (-2.25, -0.9) -- (-2.25, -7.75) node[below, black]{};
        \draw[dashed, blue, thick] (0.285, -0.9) -- (0.285, -7.75) node[below, black]{};
        \draw[dashed, blue, thick] (1.65, -4.75) -- (1.65, -7.75) node[below, black]{}; 
        
        \node[draw, fill=blue!20, rounded corners, align=center, below=0.3cm of current bounding box.south] (box1) 
            at (-11.1, -7.75) {$\ket{\psi_1}$};
        \node[draw, fill=blue!20, rounded corners, align=center, below=0.3cm of current bounding box.south] (box2) 
            at (-8.95, -7.75) {$\ket{\psi_2}$};
        \node[draw, fill=blue!20, rounded corners, align=center, below=0.3cm of current bounding box.south] (box3) 
            at (-6.55, -7.75) {$\ket{\psi_3}$};
        \node[draw, fill=blue!20, rounded corners, align=center, below=0.3cm of current bounding box.south] (box3) 
            at (-4.65, -7.75) {$\ket{\psi_4}$};
        \node[draw, fill=blue!20, rounded corners, align=center, below=0.3cm of current bounding box.south] (box3) 
            at (-2.23, -7.75) {$\ket{\psi_5}$};
        \node[draw, fill=blue!20, rounded corners, align=center, below=0.3cm of current bounding box.south] (box3) 
            at (0.3, -7.75) {$\ket{\psi_6}$};
        \node[draw, fill=blue!20, rounded corners, align=center, below=0.3cm of current bounding box.south] (box3) 
            at (1.7, -7.75) {$\ket{\psi_7}$};

    \end{tikzpicture}

\caption{\textbf{DQI circuit for max-LINSAT.} The circuit is marked with seven state snapshots, which correspond to the seven steps of the DQI algorithm given in Sec. \ref{sec:dqimaxxorsat}.  The circuit uses three registers referred to as the ``weight'' ($W$), ``error'' ($E$), and ``syndrome'' ($S$) registers. 
After the uncomputation in the $W$ and $E$ registers, qubits are returned back to the $\ket{0}$ state. The DQI state (\ref{dqistate2}) prepared on the $ S$ register after step 7.  
}
\label{DQI_circuit_and_registers}
\end{figure}
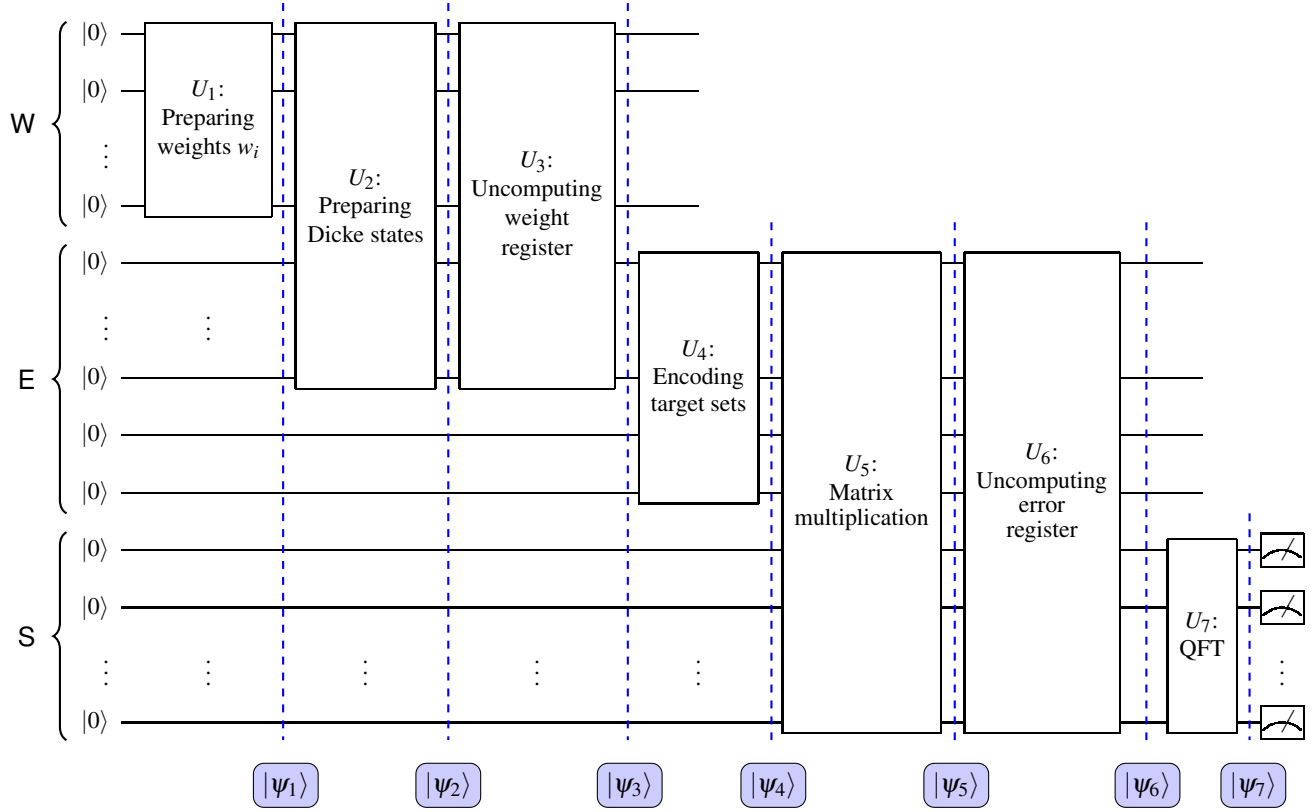

The form of the state (\ref{dqistatexorsat}) gives a hint of how the DQI algorithm proceeds.  For each of the three sums appearing in (\ref{dqistatexorsat}), there is a register where part of the state is prepared.  First, a superposition over the coefficients $ w_k $ is prepared in the ``weight'' ($W$) register.  Then Dicke states are prepared in the ``error'' ($E$) register, taking into account of the $ \vec{y} $ sum. The $W$ register is then decoupled so that the state now resides completely in the $ E$ register. Then finally, the state is transferred to the ``syndrome'' ($S$) register, where the superposition over $ \vec{x} $ is implemented. This again involves a decoupling of the $ E $ register.  A visual supplement to the steps is shown in Figure \ref{DQI_circuit_and_registers}, which gives the DQI circuit and explains its logical registers.

\paragraph{Step 1: Prepare weight factors}

The first step is to prepare the state 
\begin{align}
|\psi_1 \rangle = \sum_{k=0}^{\ell} w_k \ket{k}_W .
\end{align}
The subscript $W$ denotes that the state is stored in the $W$ register.  The $ W$ register consists of $ \lceil \log_2 (\ell +1 ) \rceil$ qubits. This can be prepared, for example, by the unitary
\begin{align}
    U_1 = e^{\frac{i \pi}{2} ( | 0 \rangle \langle \psi_1 | + | \psi_1 \rangle \langle 0 |) } ,
\end{align}
such that $ | \psi_1 \rangle = U_1 | 0 \rangle $ up to a global phase. 
Such a unitary can be implemented by quantum circuit recompilation methods \cite{cerezo2021variational}.   

By convention, when we omit registers --- in this case the $ E $ and $ S $ registers --- we assume that they are in the $ | 0 \rangle^{\otimes n} $ state, where $ n $ is the number of qubits in the register.

\paragraph{Step 2: Prepare Dicke states}

The next step is to prepare Dicke states in the $ E $ register, containing $ m $ qubits.  Dicke states are defined as
\begin{align}
|D_k^m \rangle = \frac{1}{\sqrt{\binom{m}{k}}}  \sum_{| \vec{y} | = k } | \vec{y} \rangle ,
\label{dickestate}
\end{align}
where $ \vec{y} $ is an $ m$-digit binary vector.  The restriction on the sum is such that there are $ k $ elements with a value $ 1 $ and the remaining are $ 0 $. 
The Dicke states are required in preparation for creating the symmetric polynomials, taking advantage of their similarity as shown in Table \ref{tab:sympoly}.  Methods for preparing Dicke states are well-established in the literature \cite{bacon2006efficient,bartschi2022short}.  Let us denote the unitary for preparing the Dicke state $ |D_k^m \rangle $ from the state $ | 0 \rangle^{\otimes m} $ as $ U^D_k $, which acts on the $ E $ register.  In our case we require the conditional unitary
\begin{align}
U_{2} = \sum_{k=0}^\ell |k \rangle \langle k |_W \otimes U^D_k .
\end{align}
Applying this to the state after Step 1, we have
\begin{align}
  |\psi_2 \rangle =  U_{2} | \psi_1 \rangle = \sum_{k=0}^{\ell} w_k \ket{k}_W \otimes  | D^m_k \rangle_E .
\end{align}

\paragraph{Step 3: Disentangle weight register}

Next, we disentangle the $ W $ register by uncomputing it with the unitary 
\begin{align}
    U_{3} =  \sum_{k=0}^\ell U_{3}^{(k)}  \otimes | D^m_k \rangle  \langle D^m_k |_E  ,
\end{align}
where $  U_{3}^{(k)} = e^{ \frac{\pi}{2} ( |0 \rangle \langle k | - |k \rangle \langle 0 | ) }  $ acts on the $ W $ register.  We then have the state 
\begin{align}
  |\psi_3 \rangle & = U_{3}  |\psi_2 \rangle  \nonumber \\
  & = \sum_{k=0}^{\ell} w_k | D^m_k \rangle_E 
\\
& = \sum_{k=0}^{\ell} \frac{w_k}{\sqrt{\binom{m}{k}}}  \sum_{| \vec{y} | = k } | \vec{y} \rangle_E  ,
\end{align}
where we used (\ref{dickestate}).

\paragraph{Step 4: Add constraint phases}

The next step is to imprint the phase factor $ (-1)^{\vec{v} \cdot \vec{y}} $ in (\ref{dqistatexorsat}), which has no $ \vec{x} $ dependence.  This is done by applying the unitary
\begin{align}
\label{unitary4bin}
    U_4 = Z^{v_1} \otimes Z^{v_2} \otimes \dots \otimes Z^{v_m} .
\end{align}
This results in the state
\begin{align}
\label{psi4}
 |\psi_4 \rangle & = U_4 | \psi_3 \rangle = \sum_{k=0}^{\ell} \frac{w_k}{\sqrt{\binom{m}{k}}}  \sum_{| \vec{y} | = k } (-1)^{\vec{v} \cdot \vec{y}}  | \vec{y} \rangle_E .
\end{align}

\paragraph{Step 5: Matrix multiplication}

In the next few steps we aim to create the phase factor $ (-1)^{(B^T \vec{y}) \cdot \vec{x}} $ in the DQI state.  This is performed by noting that the sum over $ \vec{x} $ in (\ref{dqistatexorsat}) is the Hadamard transform of the state $   | B^T \vec{y} \rangle $: 
\begin{align}
\label{hadamardtrans}
  H^{\otimes n}  | B^T \vec{y} \rangle = \frac{1}{\sqrt{2^n}} \sum_{\vec{x}} (-1)^{(B^T \vec{y}) \cdot \vec{x}}  | \vec{x} \rangle  .
\end{align}
Thus we aim to produce the state $  | B^T \vec{y} \rangle $ on the $S$ register, then finally take a Hadamard transform.  

The unitary that we apply for this step is 
\begin{align}
U_5 = \sum_{\vec{y}}  \sum_{\vec{x}}  | \vec{y} \rangle \langle \vec{y} |_E \otimes | \vec{x} \oplus B^T \vec{y} \rangle \langle \vec{x} |_S ,
\end{align}
which takes the same form as the oracle in the Deutsch-Jozsa algorithm \cite{deutsch1992rapid}.  Applying this, the state evolves to
\begin{align}
\label{psi5}
 |\psi_5 \rangle & = U_5 | \psi_4 \rangle = \sum_{k=0}^{\ell} \frac{w_k}{\sqrt{\binom{m}{k}}}  \sum_{| \vec{y} | = k } (-1)^{\vec{v} \cdot \vec{y}}  | \vec{y} \rangle_E  \otimes |  B^T \vec{y} \rangle_S .
\end{align}

\paragraph{Step 6: Disentangle error register}

We next wish to disentangle the $E$ register.  To achieve this, we need to apply a unitary of the form
\begin{align}
U_6 = \sum_{\vec{y}}  \sum_{\vec{x}}  | \vec{y} \oplus D_B(\vec{x}) \rangle \langle \vec{y} |_E \otimes | \vec{x} \rangle \langle \vec{x} |_S ,
\end{align}
where $ D_B(B^T \vec{x}) = \vec{x} $ inverts the map $ \vec{x} \rightarrow B^T \vec{x}  $, which is in general a non-trivial operation.  This is in fact equivalent to a decoding problem, and the complexity depends upon the nature of the $B^T $ matrix.  For matrices where the decoding can be performed efficiently, such as is the case with RS and RM codes, the unitary $ U_6 $ can be implemented efficiently.  

Applying the unitary we obtain the state
\begin{align}
    | \psi_6 \rangle = U_6 | \psi_5 \rangle = 
\sum_{k=0}^{\ell} \frac{w_k}{\sqrt{\binom{m}{k}}}  \sum_{| \vec{y} | = k } (-1)^{\vec{v} \cdot \vec{y}}  |  B^T \vec{y} \rangle_S  .
\end{align}

\paragraph{Step 7: Hadamard transform}

The final step is to apply a Hadamard transform (\ref{hadamardtrans}) to create the phase dependence $ (-1)^{(B^T \vec{y}) \cdot \vec{x}} $.  We have 
\begin{align}
    U_7 = H^{\otimes n}
\end{align}
acting on the $ S $ register to give the final state
\begin{align}
| \psi_7 \rangle = U_7 | \psi_6 \rangle =    \frac{1}{\sqrt{2^n }} \sum_{k=0}^\ell \frac{w_k}{\sqrt{\binom{m}{k}}}   \sum_{| \vec{y} | = k } \sum_{\Vec{x}}  (-1)^{ \vec{v} \cdot \vec{y} + (B^T \vec{y} ) \cdot \vec{x} }    \ket{\vec{x}}_S  = \ket{\text{DQI}} 
\end{align}

\subsection{DQI for max-LINSAT\\}
\label{sec:dqimaxlinsat}

The procedure given above for max-XORSAT provides the basic procedure of the DQI algorithm.  We now discuss the changes that are necessary when implementing the more general case for max-LINSAT. 
As described in Sec. \ref{sec:optimization}, the main difference in terms of the problem is that the vector to be optimized $ \vec{x} \in \mathbb{F}_p^n $, instead of binary variables.  Additionally, the number of elements $ r $ in the target set $ T_i $ can be greater than one.  

First, we describe how the DQI state generalizes in this case.  Following the results given in Ref. \cite{jordan2024}, this can be written
\begin{align}
\label{dqistatelinsat}
    \ket{\text{DQI}} = \ket{P(f)} = \frac{1}{\sqrt{p^n}}  \sum_{k=0}^\ell \frac{w_k}{\sqrt{\binom{m}{k}}}  
    \sum_{\substack{\vec{y} \in \mathbb{F}_p^m \\ | \vec{y} | = k} } 
    \sum_{\Vec{x}\in \mathbb{F}_p^n } 
    \left( \prod_{\substack{i=1 \\ y_i \ne 0 }}^m \tilde{g}_i (y_i) \right)     
    e^{ - \frac{2 \pi i }{p}  (B^T \vec{y} ) \cdot \vec{x} }    \ket{\vec{x}} . 
\end{align}
This has an obvious similarity to (\ref{dqistate2}), where the variables are now in $ \mathbb{F}_p $ rather than $ \mathbb{F}_2 $. In the sum over the $ \vec{y} \in \mathbb{F}_p^m $ variable, the sum is restricted to those with Hamming weight $ | \vec{y} | = k $, defined as the number of non-zero entries of the vector $ \vec{y} $.  
Here we defined the Fourier transform function as 
\begin{align}
\label{fourierg}
\Tilde{g_i}(y) = \frac{1}{\sqrt{p}} \sum_{x \in \mathbb{F}_p} e^{2 \pi i xy /p } g_i(x) ,
\end{align}
and the function 
\begin{align}
g_i(x) & := \frac{f_i(x) - \Bar{f}}{\varphi}, 
\label{gidef}
\end{align}
is a shifted and rescaled objective function for the $ i $th constraint.  The scale factors are defined as
\begin{align}
\Bar{f} &  := \frac{1}{p} \sum_{x \in \mathbb{F}_p} f_i(x) = \frac{2r - p}{p} ,\\ 
\varphi & := \sqrt{\sum_{x \in \mathbb{F}_p} |f_i(x) - \Bar{f}|^2} = \sqrt{4r\Bigl(1 - \frac{r}{p}\Bigr)} .
\end{align}
The functions above are normalized $\sum_{x \in \mathbb{F}_p}|g_i(x)|^2 = \sum_{y \in \mathbb{F}_p}|\Tilde{g_i}(y)|^2 = 1$ and have the important property that $ \tilde{g}_i (0) = 0 $. 
The DQI state (\ref{dqistatelinsat}) is derived by starting from the state (\ref{dqistate2}) but with the replacement $ f_i \rightarrow g_i $.  For the $ p = 2 $ case, $ \tilde{f}_i (y) = y \sqrt{2} (-1)^{v_i} $ and reduces to the same expression as (\ref{dqistatexorsat}). 

Due to the similarity of the state (\ref{dqistatelinsat}) to (\ref{dqistate2}), most of the steps in the preparation sequence remain the same.  The main difference is in Step 4, where the $  \tilde{g}_i (y_i) $ factors must be imprinted on the state. Here, the unitary (\ref{unitary4bin}) is replaced by
\begin{align}
U_4 = \bigotimes_{i=1}^m U_g^{(i)} ,
\end{align}
where 
\begin{align}
U_g^{(i)} = |0 \rangle \langle 0 |_E +  \sum_{y \in \mathbb{F}_p} \Tilde{g_i}(y) \ket{y} \bra{1}_E  + \dots .
\end{align}
This is a matrix where the first two columns are $ (1,0,\dots ,0 )^T $ and $ (0, \Tilde{g_i}(1), \Tilde{g_i}(2), \dots, \Tilde{g_i}(p-1))^T $, such that the state $ | 0 \rangle $ is unaffected, while the state $ | 1 \rangle $ is transformed to $ \sum_{y} \Tilde{g_i}(y) | y \rangle $.  The remaining columns (indicated by the $ \dots $), are states that are orthogonal to both $ |0 \rangle $ and  $ \sum_{y} \Tilde{g_i}(y) | y \rangle $, which are mutually orthogonal because  $  \Tilde{g_i}(0) = 0 $.  This guarantees the Hamming weight of the original Dicke state is preserved.  
The resultant state after applying this unitary then is
\begin{align}
\label{psi4linsat}
 |\psi_4 \rangle & = U_4 | \psi_3 \rangle = \sum_{k=0}^{\ell} \frac{w_k}{\sqrt{\binom{m}{k}}} \sum_{\substack{\vec{y} \in \mathbb{F}_p^m \\ | \vec{y} | = k} } 
    \left( \prod_{\substack{i=1 \\ y_i \ne 0 } }^m \tilde{g}_i (y_i) \right)  | \vec{y} \rangle_E     .
\end{align}
The product in the brackets runs over all $ i \in [1,m] $ except those where $ y_i = 0 $, since the transformation (\ref{gidef}) does not add a $ \tilde{g}_i $ function for the $ | 0 \rangle $ states in the Dicke state. 

The other change for the max-LINSAT case is that in Step 7 the Hadamard transform is generalized to a quantum Fourier transform acting on the $ S$ register
\begin{align}
    U_7 = F^{\otimes n}
\end{align}
where 
\begin{align}
    F |y \rangle =\frac{1}{\sqrt{p}} \sum_{x \in \mathbb{F}_p} e^{-2 \pi i xy/p} | x \rangle  .
\end{align}
With these replacements, the DQI state (\ref{dqistatelinsat}) is prepared.

Notice that in (\ref{dqistatelinsat}) and its preparation,  the constraint evaluation, i.e. checking which states satisfy the most constraints, is not a single explicit step as in a classical check. Instead, it is embedded in the output quantum state $\ket{\text{DQI}}$ using the constraint-dependent state preparation. This allows the quantum interference in the final QFT step to amplify the probability of measuring a good solution  $\Vec{x}$ which simultaneously satisfies a large number of different constraints. In other words, we are considering each individual constraint in parallel using the algebraic structure of correlations between constraints in order to achieve constructive quantum interference, hence the algorithm name.  This is a key strength of the DQI algorithm: it can naturally handle optimization problems where the constraints are heterogeneous (each requiring a different value or set of values), not only for homogeneous problems such as standard XOR-SAT / max-k-XORSAT.

\section{DQI performance}
\label{sec:performance}

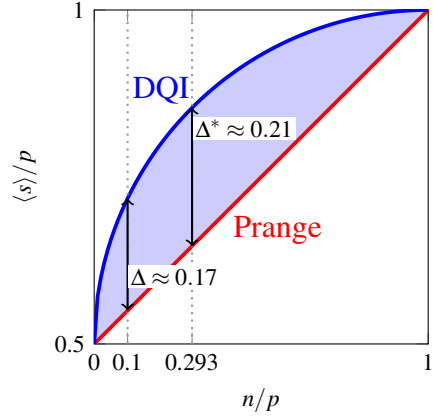
\begin{figure}[t]
    \centering
\input{Figures/performance_comparison}
    \caption{ \textbf{Performance comparison of DQI and Prange's algorithm.} 
The plot shows the expected constrain satisfaction rate $ \langle s \rangle $ versus the polynomial degree $n$, each normalized to the characteristic of the field $ p $.  
    For $n/p \approx 0.1$, we obtain the results highlighted in Fig. \ref{fig:sexp}. The largest performance gap $ \Delta $ is achieved for $n/p\approx 0.293$, where $\langle s_{\text{DQI}} \rangle/p \approx 0.854  $ and $ \langle s_{\text{Prange}} \rangle/p \approx 0.646$.     This plot is adapted from Ref. \cite{jordan2024}. }
    \label{fig:sexp2}
\end{figure}

The performance of the DQI algorithm strongly depends on the choice of weights $w_k $ in the amplifying polynomial $P$ for the DQI state (\ref{dqistate}). In Ref.  \cite{jordan2024} a procedure is introduced to optimize these weights, which maximizes the expected number of satisfied constraints $\langle s\rangle$. This helps to analytically express the asymptotic performance of the DQI algorithm. For a general max-LINSAT instance with parameters $\ell $, $m$, $p$ and $r$, the optimal choice of weights $w_k$ yields the \textit{semicircle law} \cite{jordan2024, marwaha2025DQI_complexity} 
\begin{align}
    \frac{\langle s\rangle}{m} = \left(\sqrt{\frac{\ell}{m} \left( 1 - \frac{r}{p}\right)} + \sqrt{\frac{r}{p} \left(1 - \frac{\ell}{m}\right)} \right)^2 .
\end{align}
This expression holds when $\frac{r}{p} < 1-\frac{\ell}{m}$ (non-trivial instance), while otherwise $\langle s\rangle/m = 1$. 
In Refs. \cite{jordan2024, marwaha2025DQI_complexity}, DQI was compared to many known classical algorithms for the OPI problem and it was found that DQI outperforms these algorithms. The best known classical algorithm for solving general max-LINSAT is Prange's algorithm, yielding a significantly lower constraint satisfaction rate. For example, for input parameters $n \approx \frac{p}{10}$, $r\approx \lfloor \frac{p}{2}\rfloor$, DQI achieves $\langle s_{\text{DQI}} \rangle / m  = \tfrac{1}{2}+\sqrt{\frac{n}{2p}(1-\frac{n}{p}) }\approx 0.7179$, while Prange's algorithm attains $\langle s_{\text{Prange}} \rangle/m =  \tfrac{1}{2} + \tfrac{n}{2p}   \approx 0.55$. Figure \ref{fig:sexp2} shows the performance gap between DQI and Prange's algorithm for the OPI problem.

The ability to solve the decoding problem in Step 6 is the key requirement for the efficient execution of the DQI algorithm. This requires a matrix $B$, or equivalently a linear code $C$, for which there exists an efficient decoding procedure. RS (and RM) codes are convenient examples because they admit efficient classical decoders (e.g. Berlekamp-Massey) that can correct up to half the minimum distance.

DQI exploits the fundamental mathematical property of the quantum Fourier transform for linear structures: the QFT of a given superposition over a linear subspace $C$ yields a superposition over its dual subspace $C^\perp$. Concretely, if we prepare a quantum state as a weighted superposition of elements in the solution space $C$ that encodes the objective function, applying the Fourier transform produces a state with amplitudes labeled by elements of $C^\perp$ in the syndrome space. Intuitively, in the case of DQI, the QFT effectively maps the optimization objective to syndromes. More generally, this duality between a code and its orthogonal complement is governed by the Poisson Summation Formula, which holds for any linear code over $\mathbb{F}_p^n$, not just RS codes. Hence, it is expected that DQI can be applied to \textit{any linear code}. We further discuss this connection in Appendix \ref{app:fourier}. Recent work shows that for Hermitian codes there exists a dual optimization problem called the Hermitian OPI problem, which is also efficiently solved with DQI \cite{GuJordan2025}.

\section{DQI Extensions}
\label{DQI_extensions}

There exist further improvements, generalizations, and extensions of the original DQI algorithm. Here we briefly introduce two of them.

\subsection{Multivariate OPI\\}

A generalization of the OPI problem is the multivariate OPI problem (mOPI). This is defined as follows.  The
    mOPI problem is to find a multivariate polynomial $ Q (y_1, y_2, \dots y_m ) $ with a total degree at most $u$ that maximizes 
\begin{align}
\max_{Q} | \{ \vec{y} = (y_1, y_2, \dots y_m) : Q (y_1, y_2, \dots y_m ) \in T_{(y_1, y_2, \dots y_m )} \} |,
\end{align}
where $ y_j \in \mathbb{F}_p $ is the $ j$th polynomial variable.  Here $\mathbb{F}_p$ is a finite field, where $p$ is prime and the number of variables is $m $ so that $ j \in [1,m] $.  The target sets $T_{(y_1, y_2, \dots y_m )}   \in \mathbb{F}_p $ each have a size $r \in [1, p-1] $.  The degree of the polynomial is $u \in [1, m(p-1)]$.  As with OPI, matrix $B$ is a Vandermonde matrix in the form of a Reed-Muller code.  See Example \ref{example:mOPI_instance} for a bivariate mOPI instance.

In Ref. \cite{jordan2024} it was shown that the DQI algorithm is capable of efficiently solving the mOPI problem. Significantly, mOPI contains exponentially more constraints than the univariate OPI case: $p^m$. This hints that DQI might provide an even greater quantum speedup for mOPI.


\subsection{Hamiltonian Decoded Quantum Interferometry\\}

The Hamiltonian Decoded Quantum Interferometry (HDQI) algorithm \cite{schmidhuber2025HDQI} extends the quantum decoding framework of DQI from classical diagonal Hamiltonians to general Pauli Hamiltonians.  Introduced by Schmidhuber, Poremba, Quek, and co-workers, this method reduces the Hamiltonian optimization problem (e.g. ground state diagonalization) to a problem of decoding a classical linear code associated with the Hamiltonian. In contrast to DQI which reduces classical, combinatorial optimization problems (encoded in diagonal Hamiltonians) to a decoding problem, HDQI reduces \textit{quantum optimization problems} (problems defined on more general non-diagonal Hamiltonians, where each local term $H_j$ in $H = \sum_j H_j$ acts as a quantum constraint) to a classical decoding problem.

Specifically, HDQI considers Hamiltonians of the form
\[
H = \sum_{i=1}^m v_i {\cal P}_i, \qquad v_i \in \{\pm 1\}, \quad {\cal P}_i \in \{I,X,Y,Z\}^{\otimes n},
\]
i.e., signed sums of Pauli operators. For such a Hamiltonian, the algorithm prepares the filtered state
\[
\rho(H) = \frac{P^2(H)}{\operatorname{Tr}[P^2(H)]},
\]
where $P(H)$ is the matrix obtained by applying a univariate polynomial $P(x)$ to $H$. By choosing $P$ appropriately, this construction yields Gibbs states, approximate ground states, or microcanonical ensembles.

The algorithm proceeds in three steps: (i) preparation of a ``pilot state'' on an auxiliary register that encodes the non-commutative expansion of $P(H)$; (ii) controlled applications of the Pauli terms ${\cal P}_i$ to one half of a maximally entangled state; and (iii) uncomputation of the auxiliary register via a measurement that amounts to solving a classical syndrome decoding problem. The code underlying this decoding step---called the \textit{symplectic code} of the Hamiltonian---is defined by the binary symplectic form of the Pauli operators, which captures their commutation and anticommutation structure. Interestingly, when $H$ is local, the resulting symplectic code is a classical low-density parity-check (LDPC) code, enabling efficient and highly parallelizable classical decoding.

The authors of Ref. \cite{schmidhuber2025HDQI} argue that the set of quantum optimization problems accessible to HDQI is a strict superset of those solvable by DQI: while DQI is restricted to diagonal Hamiltonians, HDQI handles non-diagonal Pauli Hamiltonians.  A subsequent work \cite{bu2026HDQI} further extends the HDQI framework to an even broader class of Hamiltonians, broadening its applicability.


\subsection{Other developments\\}

Here we briefly describe other developments relating to DQI.  

Several attempts at applying DQI to real-world  applications have been made. DQI was used in industrial-based Integer Linear Programs \cite{sabater2025BMW_DQI} and in quantum chemistry max-cut problem \cite{ralli2025QChem}, although the advantage of the latter paper was contested in a subsequent paper \cite{parekh2025noQuantumAdvantageMaxCut}.
Bu, Li, and co-workers considered the performance of DQI under noise \cite{bu2025DQIUnderNoise}, i.e. in current NISQ machines, with focus on local depolarizing noise. They found that the performance is governed by a noise-weighted sparsity parameter, with the DQI solution quality degrading exponentially with sparsity.

\begin{example}{Bivariate mOPI instance over $\mathbb{F}_{5}$}{mOPI_instance}
Let $p = 5$ and work over the finite field $\mathbb{F}_{5}$.
We consider bivariate polynomials $Q(x,y) \in \mathbb{F}_{5} $ of total degree at most $u=1$. We fix $p^m = 25$ constraints indexed by all points $(x,y)\in\mathbb{F}_5^2$,
ordered lexicographically by $y$ and then $x$. Each target set has size $r = |T_{(x,y)}| = 2$.  Concretely, we choose the constraints to be $ T_{(x,y)} = \{ v_1, v_2 \} $, where $ v_1 = 3 (x^2 + x^3 + x^4) +y $ and $ v_2 = v_1 + 1 $. The aim is to maximize the function
    \begin{align}
f(Q) = |\{(x,y) \in \mathbb{F}_p^m : Q(x,y) \in T_{(x,y)} \}| - |\{(x,y) \in \mathbb{F}_p^m : Q(x,y) \notin T_{(x,y)} \}|. 
    \end{align}

Taking the example of the candidate polynomial
\[
Q(x,y) = x + y \pmod{5}.
\]
The following table shows the target sets and the evaluation of our polynomial for each entry, i.e. $Q (x,y)\stackrel{?}{\in}T_{(x,y)}$.

{
\centering
\setlength{\arraycolsep}{3pt}
\thinmuskip=3mu
\medmuskip=4mu
\thickmuskip=5mu

\[
\begin{array}{|c|c|c|c|c|c|}
\hline
& x = 0 & x = 1 & x = 2 & x = 3 & x = 4 \\ \hline
 y = 0 & Q(0,0)=0 \in \{ 0,1\} & 1  \notin \{4,0\} & 2 \notin \{4,0\} & 3 \notin \{1,2\} &  4 \in \{3,4\} \\
y = 1 &Q(0,1) = 1  \in \{1,2\} &  2 \notin \{0,1\} & 3 \notin \{0,1\} &  4 \notin \{2,3\} &  0 \in \{4,0 \} \\
y = 2 &Q(0,2) = 2  \in \{2,3\} &  3 \notin \{1,2\} &  4 \notin \{1,2\} & 0 \notin \{3,4\} & 1 \in \{0,1\} \\
y = 3 &Q(0,3) =3  \in \{3,4\} & 4 \notin \{2,3\} & 0 \notin \{2,3\} & 1 \notin \{4,0\} & 2 \in \{1,2\} \\
y = 4 &Q(0,4) =4  \in \{4,0\} & 0 \notin \{3,4\} & 1 \notin \{3,4\} & 2 \notin \{0,1\} & 3 \in \{2,3 \} \\
\hline
\end{array}
\]
}

In this case $ Q(x_i,y_i)\in T_{(x,y)} $ in 10 of the 25 cases (the $ x=0,4 $ columns), so $f (Q) = 10 -15 = -5 $.

\begin{figure}[H]
    \centering
    \resizebox{0.7\linewidth}{!}{\input{Figures/mOPI}}
    \caption{
    \textbf{A bivariate mOPI instance over $\mathbb{F}_5$.}
    For intuition and visualization purposes, the blue surface shows the continuous version of the polynomial $Q(x,y)=x+y \pmod 5$.
    Each cuboid represents one target set (constraint) $T_{(x,y)}$ of size $r=2$.
    Green cuboids correspond to satisfied, red cuboids to unsatisfied constraints, and yellow circles represent the evaluation points for which the polynomial satisfies the constraints.
    }
    \label{fig:mOPI-2var}
\end{figure}
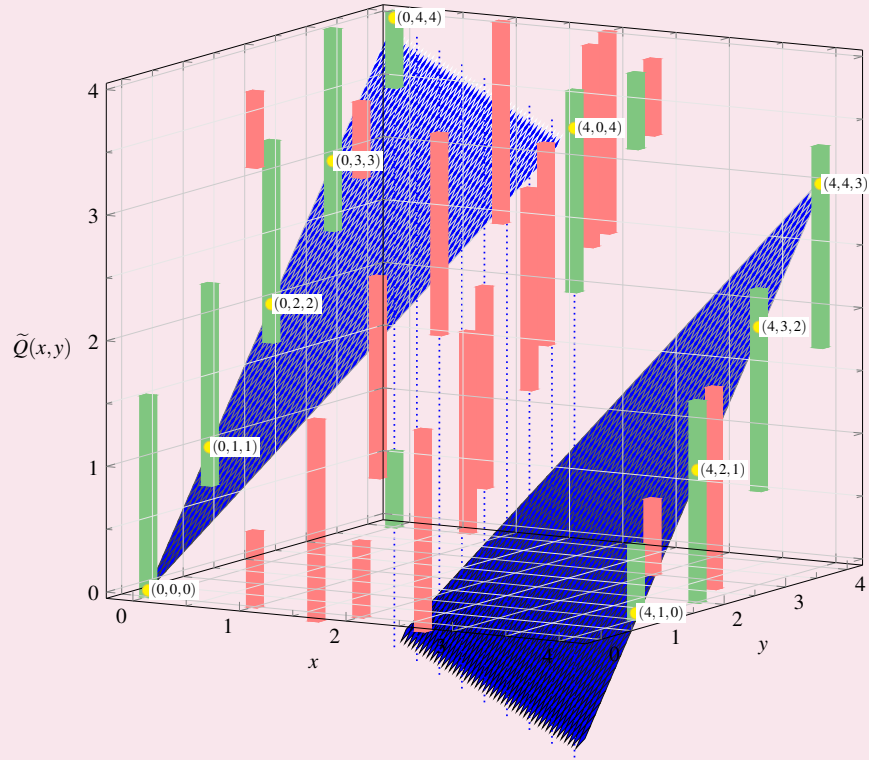

\end{example}

Several works have extended the theoretical understanding of DQI. Recently, Khattar, Jordan, and co-workers \cite{khattar2025verifiableQuantumAdvantageDQI} further improved key bottlenecks of DQI (efficient reversible decoders, in‐place architectures etc.) and made a case for it being a first known candidate for a practical, verifiable quantum advantage, with optimal asymptotic speedup. The work also suggests broader applicability: their techniques are claimed to be broadly useful beyond just OPI, e.g., Shor’s algorithm.
Chailloux and Tillich further developed DQI and also proposed a more general quantum decoding algorithm using Regev's reduction and soft decoders \cite{c_t_soft} to recover codewords from a superposition of its noisy versions. In a follow-up paper \cite{BlanvillainChaillouxTillich2025_QDP} with Blanvillain, they also extended these ideas to memory-less noise models.
Another recently published algorithm following the quantum decoding blueprint is in the realm of multivariate polynomial systems \cite{briaud2025quantumAdvantageMultivariatePolynomials}, by Briaud, Sahai and co-workers. The authors focus on the search problem of finding a solution to a system of multivariate polynomial equations, positioning it as a hard average-case NP problem.   The authors claim to provide the first evidence challenging the widespread belief that multivariate polynomial systems --- considered classically hard --- are also inherently hard for quantum computers. They show that for degrees $d \geq 3$, a structured system can be solved efficiently with a quantum algorithm, even if it remains hard classically.

\section{Summary and Outlook}
\label{Discussion_future_work}

We have presented a simplified introduction of the emerging framework of quantum decoding algorithms, with a primary focus on the DQI algorithm. The aim of the DQI algorithm is to solve the class of max-LINSAT optimization problems, including its special cases max-XORSAT and OPI. The primary technique that is used to solve the optimization problem is by polynomial amplification, where the amplitude of candidate solutions in a quantum superposition are amplified by applying a suitable polynomial.  The DQI algorithm then proceeds by constructing this quantum state.  During its construction, a disetangling procedure is required, which requires decoding a dual linear code. For the OPI problem, the dual problem is the problem of decoding a Reed-Solomon code, for which efficient solutions are known to exist. 
The algorithm exploits the Fourier duality between the solution space and its syndrome space, allowing for the optimization of heterogeneous constraints to be performed in parallel.  The performance of DQI, quantified by the semicircle law, demonstrates a provably higher expected constraint satisfaction rate compared to the best known classical algorithms such as Prange's algorithm for certain parameter regimes. Some important extensions of the original DQI work, including its multivariate generalization (mOPI) and the more recent Hamiltonian Decoded Quantum Interferometry (HDQI) algorithm, which broadens the scope from diagonal Hamiltonians to general Pauli Hamiltonians.

There is strong evidence that DQI possesses a superpolynomial advantage over equivalent classical methods in approximately solving OPI \cite{jordan2024}.  There is, however, no statement of a general quantum speedup in solving the more general max-LINSAT problem.  The reason is that the decoding step in the DQI algorithm needs to be performed efficiently, and this is not always available.  Thus the superpolynomial improvement is limited to subproblems such as OPI where an efficient classical decoder is available.  There is some evidence that the OPI problem is classically hard to simulate \cite{marwaha2025DQI_complexity}.   For the max-XORSAT case, there are fast classical algorithms that can match or beat DQI, hence there is no quantum speedup in these cases.  The case that does have a proven quantum advantage is in terms of query complexity, in the same way as shown by Yamakawa and Zhandry \cite{YZ22}.  This is however not a clear complexity improvement of solving a particular optimization problem, but reveals the nature of efficient solution of the DQI algorithm.

There are multiple interesting open questions.  The immediate question is whether the method can be applied to other classes of optimization problems. Broadening the scope by mapping other problems to max-LINSAT, OPI would be a welcome development. Investigations revealing  algorithmic adjustments of DQI to further improve its performance are underway \cite{c_t_soft}.  Another important question is whether it is possible to characterize the hardness of problems tackled by quantum decoding algorithms and relate it to the complexity of decoding problems (e.g., when are they NP-adjacent, when do they lie in BQP, etc.).  Some more specific questions remain worthy of investigation.  For example, typically the size of set of conditions $|T_i| = r$ is kept constant. Is there a  DQI performance change when this is relaxed? Finally, two promising methods for classical simulation of DQI remain largely unexplored: neural quantum states  and tensor networks. Although recent complexity analysis \cite{marwaha2025DQI_complexity} suggests that DQI may be classically hard to simulate, this question warrants further systematic investigation using these approaches.

In the context of quantum computer applications of optimization problems, the broader implications of the DQI technique appear to be promising.  While it has long been suggested that quantum computers may be useful for optimization problems, establishing provable speedups has remained challenging. To date, reported speedups have primarily been demonstrated on carefully constructed or structured benchmark instances \cite{mohseni2022ising}.  
We note that there have recently been rapid advances showing quantum speedups in several classes of optimization problems.  For example, evidence for an exponential speedup for max-cut of high girth 3-regular graphs using QAOA was shown \cite{farhi2025lower}.  In Ref. \cite{herman2025mechanisms}, polynomial-time optimization of nonconvex functions using the real-space adiabatic algorithm was proven.  Hence quantum optimization remains a rapidly developing field, with DQI being one of several exciting developments.  Continued investigation of quantum optimization algorithms, including quantum decoding approaches,  may provide a pathway toward demonstrating quantum advantage on practically relevant optimization problems.

\bibliography{references}

\section*{Acknowledgements}
We thank Stephen Jordan for illuminating discussions and helpful feedback on draft of this tutorial.  This work is supported by the SMEC Scientific Research Innovation Project (2023ZKZD55); the Science and Technology Commission of Shanghai Municipality (22ZR1444600); the NYU Shanghai Boost Fund; the China Foreign Experts Program (G2021013002L); the NYU-ECNU Institute of Physics at NYU Shanghai; the NYU Shanghai Major-Grants Seed Fund; and Tamkeen under the NYU Abu Dhabi Research Institute grant CG008; SRPP program at the NYU Shanghai.




\appendix
\renewcommand{\thesubsection}{\thesection.\Roman{subsection}}
\setcounter{equation}{0}
\renewcommand{\theequation}{A.\arabic{equation}}

\section{Symmetric polynomials}
\label{Elementary_symmetric_polynomials}

In this section, we discuss further details regarding the symmetric polynomials and their associated state. 

First, let us define the state associated to the symmetric polynomial
\begin{align}
   \ket{P^{(k)}} \propto  \sum_{\vec{x} \in \mathbb{F}^n_p} P^{(k)} (g_1, g_2, \dots, g_m) | \vec{x} \rangle ,
\end{align}
where we have written a proportionality since we have omitted normalization constants. We have used the symmetric polynomial for the max-LINSAT version of the DQI state, where the replacement $ f_i \rightarrow g_i $ has been made in (\ref{dqistate2}).  Its quantum Fourier transformed state is
\begin{align}
   \ket{\tilde{P}^{(k)}} = (F^{-1})^{\otimes n} \ket{P^{(k)}} \propto  \sum_{\vec{x} \in \mathbb{F}^n_p} P^{(k)} (\tilde{g}_1, \tilde{g}_2, \dots, \tilde{g}_m) | \vec{x} \rangle  ,
\end{align}
where the $ \tilde{g}_i $ are defined in (\ref{fourierg}).  These states form an orthonormal set $\{\ket{P^{(k)}}\}$, $\{\ket{\Tilde{P}^{(k)}}\}$, provided that $\ell < \frac{d^\perp}{2}$, where $ d^\perp $ is the minimum distance of the dual code $ C^\perp $.  This stems from the fact that for the set of $\Vec{y} \in \mathbb{F}_p^m$ with the minimal distance $d^\perp$ assumption, corresponding vectors $B^T\Vec{y}$ are all distinct. This enables us to express the DQI state $\displaystyle \ket{P(f)} = \sum_{k=0}^\ell w_k \ket{P^{(k)}}$ and its dual $\displaystyle \ket{\Tilde{P}(f)} = \sum_{k=0}^\ell w_k \ket{\Tilde{P}^{(k)}}$ in these bases. This is significant because orthonormality ensures that these different error patterns are perfectly distinguishable in the syndrome register. There is no unwanted overlap between different errors $\Vec{y}$ and $\Vec{y}'$. This property also simplifies the algorithm's performance guarantees, as we are able to precisely tune the algorithm's hyperparameters --- the weights $\Vec{w}$.

Another way to interpret these symmetric polynomials comes from Fourier analysis. In fact, the $P^{(k)}$ can be interpreted as spectral filters of the Fourier spectrum of the function $P(f(\vec{x}))$. In other words, they are functions modifying the frequency spectrum of the original function, as they isolate a specific part of that spectrum. We can see that $P^{(k)}$ decompose $P(f)$ into so-called frequency bands in syndrome space.

Applying a quantum Fourier transform to the solution space state $\ket{P^{(k)}}$ maps its input $\Vec{x}$ to the space supported by the 
syndrome vectors $\Vec{s} = B\Vec{y}$. This is analogous to the conventional interpretation of the Fourier transform, where the \textit{time domain} is the space of solutions $\Vec{x}$ --- polynomials we are trying to optimize over --- and the \textit{frequency domain} is the space of syndromes $\Vec{s}$ --- the linear combinations of constraints. The transformed DQI state 
\begin{align}
\ket{\tilde{P}(f)} = (F^{-1})^{\otimes n} \ket{P (f)} = \sum_{k=0}^\ell w_k \underbrace{\frac{1}{\sqrt{\binom{m}{k}}}  
    \sum_{\substack{\vec{y} \in \mathbb{F}_p^m \\ | \vec{y} | = k} } 
    \left( \prod_{\substack{i=1 \\ y_i \ne 0 }}^m \tilde{g}_i (y_i) \right) \ket{B^T \vec{y} } }_{\text{$k$th frequency band}} . 
\end{align}
is a linear combination of up to $\ell$ frequency bands in the Fourier spectrum.  Each frequency band carries information about the correlations among $k$-tuples of constraints, with weights $w_k$ controlling how much each band contributes to the final state. We note a common misconception is that if the coefficient $w_j$ of the $j$th frequency band is the largest, then $P(f)$ satisfies $j$ constraints. There is no such direct correlation.

Finally, we clarify what the role of the precomputed weights is. Weights $\Vec{w}$ are not direct embeddings of our problem data. They are internal parameters of the DQI algorithm chosen to optimize its performance, i.e., to maximize the expected number of satisfied constraints. While we can express $P(f)$ using symmetric polynomials $\displaystyle P(f)=\sum_{k=0}^\ell \frac{w_k}{\sqrt{p^n\binom{m}{k}}} P^{(k)}$, we cannot evaluate any constraints without the matrix $B$; this only represents the distribution over all $P^{(k)}$, but not the values of those $P^{(k)}$. In a classical algorithm, computing this amplifying polynomial $P(f)$ would not help evaluate candidate solutions $\Vec{x}$, but in the quantum setting, the interference patterns created by these weighted combinations enable the amplification of high-quality solutions.


\begin{example}{Fourier transform relationship between code and syndrome space}{FT_codes}
Let $G_1$ be the generating matrix of code $C$:
\setlength{\belowdisplayskip}{0pt}
\[
G_1 = \begingroup\renewcommand{\arraystretch}{0.5}
    \begin{pmatrix}
    1 & 0 & 1 & 1 \\
    0 & 1 & 1 & 2 
    \end{pmatrix}
    \endgroup .
\]

The code $C$ can be understood as the row space of $G_1$: $C = \operatorname{span}\{(1,0,1,1), (0,1,1,2)\}$, a 2-dimensional subspace of $\mathbb{F}_5^4$. Its dual code is $C^\perp = \{\vec{d} \in \mathbb{F}^4_5: G_1 \vec{d}^T = 0\}$, where solving $G_1 \vec{d}^T = 0$ is equivalent to $\left\{
\begin{array}{@{}l@{}}
d_1 + d_3 + d_4 = 0 \\
d_2 + d_3 + 2d_4 = 0
\end{array}
\right. .$ \\

A basis for $C^\perp$ is hence given by the rows of
\(
G_2 = \begingroup\renewcommand{\arraystretch}{0.5}
    \begin{bmatrix}
    4 & 4 & 1 & 0 \\
    4 & 3 & 0 & 1 
    \end{bmatrix}
    \endgroup
\), or in systematic form [$\mathbb{I}_2|\mathbb{P}$]:
\(
    \begingroup\renewcommand{\arraystretch}{0.5}
    \begin{bmatrix}
    1 & 0 & 3 & 1 \\
    0 & 1 & 1 & 4 
    \end{bmatrix}
    \endgroup
\).
To verify orthogonality, we multiply $G_1$ and $G_2^T$:
\[
G_1 G_2^T = \begingroup\renewcommand{\arraystretch}{0.5}\begin{bmatrix}
1 & 0 & 1 & 1 \\
0 & 1 & 1 & 2 
\end{bmatrix}
\begin{bmatrix}
1 & 0 \\
0 & 1 \\
3 & 1 \\
1 & 4 
\end{bmatrix} = \begin{bmatrix}
5 & 5 \\
5 & 10 
\end{bmatrix} \equiv \begin{bmatrix}
0 & 0 \\
0 & 0 
\end{bmatrix} \pmod{5}.
\endgroup
\] \\

Next, we show the Fourier transform relationship between $C$ and $C^\perp$. Let $\displaystyle \ket{C} = \frac{1}{\sqrt{|C|}} \sum_{\Vec{c} \in C} \ket{\vec{c}}$ and $\displaystyle \ket{C^\perp} = \frac{1}{\sqrt{|C^\perp|}} \sum_{\Vec{d} \in C^\perp} \ket{\vec{d}}$ be uniform superpositions of quantum states over $C$ and $C^\perp$, respectively. The QFT over $\mathbb{F}_5^4$ is defined as 
\begin{align}
\displaystyle F\ket{\vec{y}} = \frac{1}{5^2} \sum_{\Vec{x} \in \mathbb{F}_5^4} e^\frac{2\pi i \Vec{x}\cdot \Vec{y} }{5}  \ket{\vec{x}}.
\end{align}

This claim can be reiterated as: $F\ket{C} = \ket{C^\perp}$ and $F\ket{C^\perp} = \ket{C}$.

Since $C$ is a linear code, we can apply QFT to every component of the superposition:
\begin{align}
    F\ket{C} &= \frac{1}{\sqrt{|C|}} \sum_{\Vec{c} \in C} F\ket{\Vec{c} } = \frac{1}{5^2\sqrt{|C|}} \sum_{\Vec{x} \in \mathbb{F}_5^4} \sum_{\Vec{c} \in C} e^{2 \pi i \Vec{x} \cdot \Vec{c}/5}  \ket{\Vec{x} } \\
    &= \frac{1}{25\sqrt{|C|}} |C| \sum_{\Vec{x} \in C^\perp} \ket{\Vec{x} } = \frac{\sqrt{|C|}}{25} \sum_{\Vec{x} \in C^\perp} \ket{\Vec{x}}
\end{align}
Here we used the orthogonality property: if $\Vec{x} \in C^\perp$, then $\Vec{x} \cdot \Vec{c} = 0$ for all $\Vec{c}$, so the inner sum becomes $\displaystyle \sum_{\Vec{c} \in C} 1 = |C|$. If $\Vec{x} \notin C^\perp$, destructive interference makes the sum $0$.

Finally, we use the identity $|C|  |C^\perp| = 5^2 \times 5^2 = 625$, so $ \displaystyle \frac{\sqrt{|C|}}{25} = \frac{5}{25} = \frac{1}{5} = \frac{1}{\sqrt{|C^\perp|}} $. Thus, we arrive at the definition of the dual state:
\begin{align}
    \frac{1}{\sqrt{|C^\perp|}} \sum_{\Vec{x} \in C^\perp} \ket{\Vec{x} } = \ket{C^\perp}.
\end{align}
\end{example}

\section{The Fourier transform in coding theory}
\label{app:fourier}

In the DQI approach, it is implied that the solution space and syndrome space are connected via the Fourier transform. Concretely, the algorithm exploits the fact that codewords from $C$ and syndromes from $C^\perp$ are connected through the Fourier transform to allow efficient sampling from all of $\mathbb{F}^n_p$. This property holds not only for Reed-Solomon codes, but for all linear codes. It relies on the fundamental identity for linear subspaces: $|C|  |C^\perp| = p^n$. This behavior is formally described by the following well-known theorem \cite{terras1999fourier} in coding theory \footnote{This formulation of the theorem is a vector space adaptation of the group theory formulation of the original theorem.}, which we state without proof.

\begin{theorem}{Poisson Summation Formula for Linear Codes\\}{}
Let $C \subseteq \mathbb{F}_p^n$ be a linear code (a linear subspace). Let $\ket{C}$ be the uniform superposition state over the codewords of $C$:$$ \ket{C} = \frac{1}{\sqrt{|C|}} \sum_{\vec{c} \in C} \ket{\vec{c}} $$The Quantum Fourier Transform maps $\ket{C}$ to the uniform superposition of the dual code $C^\perp$
$$ F \ket{C} = \ket{C^\perp} = \frac{1}{\sqrt{|C^\perp|}} \sum_{\vec{c}^\perp \in C^\perp} \ket{\vec{c}^\perp}. $$
\end{theorem}

Following standard coding theory notation, let us denote $C \subseteq \mathbb{F}^n_p$ (solution space) and its dual $C^\perp \subseteq \mathbb{F}^n_p$ (syndrome space); both subspaces are of the same ambient space of dimension $n$. They are connected through the $k \times n$ generator matrix $G$, where $C$ is the row space of $G$, and the dual code $C^\perp$ is the kernel of $G$. We can see this structure in Example \ref{example:FT_codes} with a simple linear code and its dual over $\mathbb{F}_5$.

\end{document}

%% file: Figures/performance_comparison.tex
\begin{tikzpicture}
\begin{axis}[
    width=6cm, height=6cm,
    axis lines=box,
    xmin=0, xmax=1,
    ymin=0.5, ymax=1,
    xtick={0, 0.1, 0.293, 1},
    xticklabels={0, 0.1, 0.293, 1},
    ytick={0.5, 1},
    yticklabels={0.5, 1},
    xlabel={$n/p$},
    ylabel={$\langle s \rangle/p$},
    tick label style={font=\small},
    label style={font=\small},
    every axis plot post/.append style={mark=none}
]

\addplot[red, line width = 1.4pt, name path=prange] coordinates {(0, 0.5) (1, 1)};
\addplot[blue, line width = 1.4pt, name path=dqi, domain=0:1, samples=100] {0.5 + 0.5*sqrt(2*x - x^2)};

\addplot[blue!20] fill between[of=prange and dqi];

\draw[dotted, thick, black!40] (axis cs:0.1, 0.5) -- (axis cs:0.1, 1);
\draw[dotted, thick, black!40] (axis cs:0.293, 0.5) -- (axis cs:0.293, 1);
\draw[<->, thick, black] 
    (axis cs:0.293, 0.646) -- (axis cs:0.293, 0.854)
    node[pos=0.85, right, font=\small, fill=white, inner sep=1pt] {$\Delta^* \approx 0.21$};
\draw[<->, thick, black] 
    (axis cs:0.1, 0.55) -- (axis cs:0.1, 0.718)
    node[pos=0.3, right, font=\small, fill=white, inner sep=1pt] {$\Delta \approx 0.17$};

\node[blue, font=\large] at (axis cs:0.2, 0.88) {DQI};
\node[red, font=\large] at (axis cs:0.55, 0.67) {Prange};

\end{axis}
\end{tikzpicture}

%% file: Figures/mOPI.tex
\begin{tikzpicture}
\begin{axis}[
    view={30}{10},
    width=0.85\linewidth,
    xlabel={$x$},
    ylabel={$y$},
    zlabel={$\widetilde{Q}(x,y)$},
    zlabel style={rotate=-90}, 
    xmin=-0.25,
    xmax=4.25,
    ymin=-0.25,
    ymax=4.25,
    zmin=-0.05,
    zmax=4.05,
    domain=0:4,
    y domain=0:4,
    samples=50,
    samples y=50,
    xtick={0,1,2,3,4},
    ytick={0,1,2,3,4},
    ztick={0,1,2,3,4},
    grid=both,
    major grid style={draw=gray!40, line width=0.4pt, opacity=0.4},
    minor grid style={draw=gray!20, line width=0.2pt, opacity=0.7},
    minor tick num=1,
    axis on top,
    clip=false,
    colormap={mygray}{
        color(0cm)=(black);
        color(1cm)=(white)
    },
]

\addplot3[
    surf,
    draw=none,
    fill=blue, 
    shader=flat,
    opacity=0.7, 
    domain=0:4,
    y domain=0:4,
    restrict expr to domain={x+y}{0:3.999}
] {x+y};

\addplot3[
    surf,
    draw=none,
    fill=blue, 
    shader=flat,
    opacity=0.7, 
    domain=0:4.05,
    y domain=0:4.05,
    restrict expr to domain={x+y}{4.001:8}
] {x+y-5};

\foreach \x in {0, 0.5, 1, 1.5, 2, 2.5, 3, 3.5, 4} {
    \addplot3[
        blue, 
        dotted, 
        thick, 
        line join=round,
        samples=2,
        samples y=1,
    ] coordinates {(\x, {4-\x}, -1.01) (\x, {4-\x}, 3.99)};
}
\newcommand{\cuboidC}[7]{%

\pgfmathsetmacro{\xa}{#1-#4}
\pgfmathsetmacro{\xb}{#1+#4}
\pgfmathsetmacro{\ya}{#2-#5}
\pgfmathsetmacro{\yb}{#2+#5}
\pgfmathsetmacro{\za}{#3-#6}
\pgfmathsetmacro{\zb}{#3+#6}

\addplot3[fill=#7, draw=#7, opacity=0.35]
coordinates {
(\xa,\ya,\za) (\xb,\ya,\za) (\xb,\yb,\za) (\xa,\yb,\za)
} -- cycle;

\addplot3[fill=#7, draw=#7, opacity=0.35]
coordinates {
(\xa,\ya,\zb) (\xb,\ya,\zb) (\xb,\yb,\zb) (\xa,\yb,\zb)
} -- cycle;

\addplot3[fill=#7, draw=#7, opacity=0.35]
coordinates {
(\xa,\ya,\za) (\xb,\ya,\za) (\xb,\ya,\zb) (\xa,\ya,\zb)
} -- cycle;

\addplot3[fill=#7, draw=#7, opacity=0.35]
coordinates {
(\xa,\yb,\za) (\xb,\yb,\za) (\xb,\yb,\zb) (\xa,\yb,\zb)
} -- cycle;

\addplot3[fill=#7, draw=#7, opacity=0.35]
coordinates {
(\xa,\ya,\za) (\xa,\yb,\za) (\xa,\yb,\zb) (\xa,\ya,\zb)
} -- cycle;

\addplot3[fill=#7, draw=#7, opacity=0.35]
coordinates {
(\xb,\ya,\za) (\xb,\yb,\za) (\xb,\yb,\zb) (\xb,\ya,\zb)
} -- cycle;
}

\def\hit{green!55!black!50}
\def\miss{red!50}
\def\d{0.05}
\def\hWrap{0.3}   


\cuboidC{0}{0}{0.75}{\d}{\d}{0.8}{\hit}
\cuboidC{0}{1}{1.5}{\d}{\d}{0.8}{\hit}
\cuboidC{0}{2}{2.5}{\d}{\d}{0.8}{\hit}
\cuboidC{0}{3}{3.25}{\d}{\d}{0.8}{\hit}
\cuboidC{0}{4}{3.75}{\d}{\d}{\hWrap}{\hit}
\cuboidC{0}{4}{0.25}{\d}{\d}{\hWrap}{\hit}

\cuboidC{1}{0}{3.75}{\d}{\d}{\hWrap}{\miss}
\cuboidC{1}{0}{0.25}{\d}{\d}{\hWrap}{\miss}
\cuboidC{1}{1}{0.5}{\d}{\d}{0.8}{\miss}
\cuboidC{1}{2}{1.5}{\d}{\d}{0.8}{\miss}
\cuboidC{1}{3}{2.5}{\d}{\d}{0.8}{\miss}
\cuboidC{1}{4}{3.25}{\d}{\d}{0.8}{\miss}

\cuboidC{2}{0}{3.75}{\d}{\d}{\hWrap}{\miss}
\cuboidC{2}{0}{0.25}{\d}{\d}{\hWrap}{\miss}
\cuboidC{2}{1}{0.5}{\d}{\d}{0.8}{\miss}
\cuboidC{2}{2}{1.5}{\d}{\d}{0.8}{\miss}
\cuboidC{2}{3}{2.5}{\d}{\d}{0.8}{\miss}
\cuboidC{2}{4}{3.25}{\d}{\d}{0.8}{\miss}

\cuboidC{3}{0}{1.5}{\d}{\d}{0.8}{\miss}
\cuboidC{3}{1}{2.5}{\d}{\d}{0.8}{\miss}
\cuboidC{3}{2}{3.5}{\d}{\d}{0.8}{\miss}
\cuboidC{3}{3}{3.75}{\d}{\d}{\hWrap}{\miss}
\cuboidC{3}{3}{0.25}{\d}{\d}{\hWrap}{\miss}
\cuboidC{3}{4}{0.5}{\d}{\d}{0.8}{\miss}

\cuboidC{4}{0}{3.5}{\d}{\d}{0.8}{\hit}
\cuboidC{4}{1}{4}{\d}{\d}{\hWrap}{\hit}
\cuboidC{4}{1}{0.25}{\d}{\d}{\hWrap}{\hit}
\cuboidC{4}{2}{0.75}{\d}{\d}{0.8}{\hit}
\cuboidC{4}{3}{1.5}{\d}{\d}{0.8}{\hit}
\cuboidC{4}{4}{2.5}{\d}{\d}{0.8}{\hit}

\addplot3[
    only marks,
    mark=*,
    mark size=2.8pt,
    draw=yellow!80!black,
    fill=yellow,
    opacity=1.0,
    nodes near coords,
    point meta=explicit symbolic,
    nodes near coords style={
        font=\scriptsize,
        fill=white,
        draw=none,
        inner sep=1pt,
        anchor=west
    }
]
coordinates {
    (0,0,0) [\((0,0,0)\)]
    (4,0,4) [\((4,0,4)\)]

    (0,1,1) [\((0,1,1)\)]
    (4,1,0) [\((4,1,0)\)]

    (0,2,2) [\((0,2,2)\)]
    (4,2,1) [\((4,2,1)\)]

    (0,3,3) [\((0,3,3)\)]
    (4,3,2) [\((4,3,2)\)]

    (0,4,4) [\((0,4,4)\)]
    (4,4,3) [\((4,4,3)\)]
};

\end{axis}
\end{tikzpicture}